\documentclass[12pt,letterpaper]{article}
\usepackage[a4paper, total={7in, 10in}]{geometry}

\usepackage{graphicx}
\usepackage{helvet}
\usepackage{authblk}
\usepackage{hyperref}
\usepackage{amsmath} 
\usepackage{amssymb} 
\usepackage{orcidlink} 
\usepackage[super,comma,sort&compress]  
   {natbib}
\usepackage[title]{appendix}%
\usepackage{float}

\usepackage[utf8]{inputenc} 
\usepackage[T1]{fontenc}    
\usepackage{url}            
\usepackage{booktabs}       
\usepackage{amsfonts}       
\usepackage{nicefrac}       
\usepackage{microtype}      
\usepackage{xcolor}         
\usepackage{multirow}
\usepackage{threeparttable}
\usepackage{amsmath}
\usepackage{subcaption}
\usepackage{makecell}
\usepackage{placeins}
\usepackage[utf8]{inputenc} 
\usepackage{booktabs}      
\usepackage{siunitx}       
\usepackage{caption}       
\usepackage{tabularx}
\usepackage{listings}    
\usepackage{pifont}
\usepackage{makecell}
\usepackage{caption}
\usepackage{enumitem}
\usepackage{pdfpages}
\usepackage{placeins} 
\usepackage{longtable}
\usepackage{pdflscape}   
\usepackage{array}
\usepackage{rotating}
\usepackage{booktabs,multirow,adjustbox,makecell}
\theadset{\bfseries}

\makeatletter
\renewcommand{\maketitle}{\bgroup\setlength{\parindent}{0pt}
\begin{flushleft}
  \textbf{\@title}
  
  \@author
\end{flushleft}\egroup}
\makeatother

\title{Smartwatch Photoplethysmography-Derived Heart Age via ECG-Guided Cross-Modal Pretraining as a Digital Biomarker of Vascular Aging
}
\date{}

\author[1,2,3,\orcidlink{0009-0008-6988-2183}]{Donglin Xie}
\author[3]{Xueying Gui}
\author[3]{Yutian Zhu}
\author[3]{Feng Xu}
\author[1,4]{Guangkun Nie}
\author[3,5]{Chenyang Xu}
\author[1,2]{Jun Li}
\author[6]{Shuailong Tang}
\author[3]{Xiaoyu Li}
\author[3]{Qi Xie}
\author[3,*]{Yelei Li}
\author[1,2,7,*,\orcidlink{0000-0001-7521-5127}]{Shenda Hong}

\affil[1]{National Institute of Health Data Science, Peking University, Beijing, China}
\affil[2]{Institute of Medical Technology, Peking University Health Science Center, Beijing, China}
\affil[3]{OPPO Health Lab, Shenzhen, China}
\affil[4]{School of Intelligence Science and Technology, Peking University, Beijing, China}
\affil[5]{Xidian University, Xi'an, China}
\affil[6]{Shenzhen Technology University, Shenzhen, China}
\affil[7]{Institute for Artificial Intelligence, Peking University, Beijing, China}

\affil[*]{Correspondence: hongshenda@pku.edu.cn, liyelei1@oppo.com}

\begin{document}

\maketitle


\section*{ABSTRACT}
Digital biomarkers of cardiovascular aging derived from physiological signals, commonly
referred to as heart age or vascular age, have been extensively studied and shown to associate
with a range of diseases and cardiovascular outcomes. However, most existing work relies on
resting electrocardiography (ECG), imaging, or specialized clinical vascular function
measurements, and studies that link wearable photoplethysmography (PPG) to clinically
relevant phenotypes such as arterial stiffness and hypertension remain scarce, particularly
in the low-burden, repeatable setting of smartwatches. To address this gap, we developed
an ECG-guided cross-modal pretraining framework that leverages synchronized smartwatch
ECG during training to enhance PPG representation learning, while relying solely on PPG
at inference to match the intended smartwatch deployment scenario, and systematically
evaluated the resulting model-derived heart age gap in relation to arterial stiffness and
prevalent hypertension. The study included three cohorts sourced from OPPO across China, comprising
581,804 participants and 7,452,131 recordings. The Vascular Health Study (VHS) cohort was used for
synchronized ECG--PPG self-supervised pretraining, supervised fine-tuning, and internal
validation; two external cohorts, pulse wave velocity (PWV) and home blood pressure monitoring (HBPM), were used
for the corresponding arterial stiffness and hypertension prevalence association
analyses. The
model combined subject-aware self-supervised learning with ECG--PPG contrastive alignment
and was deployed using PPG-only inference. The SA-CLIP-pretrained PPG-only model achieved
subject-level MAE values of 5.895 years with Pearson's $r=0.819$ in the external PWV
cohort and 4.344 years with $r=0.800$ in the HBPM cohort. Short-term aggregation of
repeated recordings further improved subject-level stability. Heart age gap remained
significantly associated with PWV after controlling for chronological age, with a partial
correlation of 0.2627 ($P < 0.001$); multivariable OLS showed that each 1-year higher
heart age gap was associated with 0.062 m/s higher PWV. Participants with accelerated
heart aging had 0.91 m/s higher adjusted PWV than those with decelerated heart aging.
In the HBPM cohort, each 1-SD higher adjusted heart age gap was associated with higher
odds of prevalent hypertension (OR 1.72, 95\% CI 1.49--1.99), and the highest quartile
had an OR of 4.25 compared with the lowest quartile. These findings suggest that
ECG-guided pretraining enhances PPG heart-age representations and that smartwatch
PPG-derived heart age gap may serve as a scalable digital biomarker for stratifying
arterial stiffness and prevalent hypertension.

\section*{KEYWORDS}
Smartwatch, Photoplethysmography, Heart age, Cross-modal pretraining, Vascular aging


\section*{INTRODUCTION}



Cardiovascular disease remains one of the leading contributors to global morbidity and mortality, with hypertension and arterial stiffness representing key modifiable risk factors of cardiovascular events and disease progression \citep{joseph2017reducing, mensah2019global, chong2025global, vaduganathan2022global, global2025global, gheorghe2018economic, kario2024global, world2023global, zhou2021global}. Increased arterial stiffness reflects age-related changes in vascular structure and function and is closely associated with elevated blood pressure, increased left ventricular load, target-organ damage, and future cardiovascular events \citep{kim2023arterial, segers2020measure, fiori2021non, boutouyrie2021arterial, aimagambetova2024arterial, sequi2020accuracy, zhong2018carotid, chang2026glycosylation}. However, hypertension and vascular aging are often clinically silent in their early stages, and many individuals do not undergo sufficient risk assessment before overt disease or organ damage becomes apparent. Digital tools that can identify early, asymptomatic cardiovascular aging signals in daily life with low burden and repeatability may therefore have important public health value \citep{baghdadi2023advanced, karunathilake2018secondary, ullah2023smart, willis2022digital, benis2021one, zhang2026ecgomics}.

The concepts of biological age and heart age provide an intuitive framework for individualized cardiovascular risk characterization. Unlike chronological age, heart age or vascular age aims to translate complex cardiovascular status into an interpretable age scale, thereby capturing physiological deviation from age-matched expectations \citep{jylhava2017biological, li2023progress, attia2019age, lima2021deep, nie2025artificial}. Within this framework, the heart age gap, defined as the difference between predicted heart age and chronological age, can be used as a digital phenotype that quantifies accelerated or decelerated cardiovascular aging \citep{ladejobi202112, park2024artificial, cho2025artificial, hempel2025explainable, al2025advanced}. Previous studies have explored heart-age or vascular-age estimation using conventional risk factors, ECG, imaging, or clinical vascular function measurements and have suggested that such age gaps are associated with adverse cardiovascular phenotypes or outcomes \citep{raghu2021deep, grimbly2024estimating, lindow2022heart}. Nevertheless, many existing approaches rely on clinical settings, active examinations, or specialized equipment, limiting their suitability for high-frequency, continuous, and low-burden monitoring. Whether heart-age models in wearable settings can reflect clinically relevant cardiovascular phenotypes such as arterial stiffness and hypertension in external cohorts remains insufficiently validated \citep{roos2023wearable, miller2025wearable, shanmugam2026wearable}.

Smartwatch photoplethysmography provides a unique opportunity for large-scale cardiovascular aging assessment. PPG records peripheral pulse waveforms optically and can be acquired conveniently, repeatedly, and with low user burden \citep{wang2016algorithmic, almarshad2022diagnostic, niescoping, weiler2017wearable}. The PPG waveform is influenced by arterial compliance, peripheral resistance, pulse wave propagation, reflected waves, and vascular tone, and may therefore contain information related to vascular aging and blood pressure burden \citep{rodriguez2021development, ferizoli2024arterial, huotari2011photoplethysmography, shin2017feasibility}. Recent studies using PPG or wearable physiological signals for age estimation suggest that PPG contains signal patterns associated with age-related physiological changes. However, smartwatch PPG age prediction still faces several challenges, including scarcity of labeled data, signal noise and variability in wearing conditions, generalization across devices and cohorts, and insufficient physiological interpretation of model outputs. In particular, existing PPG age models have rarely been linked simultaneously to gold-standard vascular stiffness measurements and real-world blood pressure monitoring phenotypes \citep{miller2025wearable, nie2025artificial, charlton2022assessing}.

\begin{figure}
\centering
\includegraphics[
    width=\textwidth,
    height=0.75\textheight,
    keepaspectratio
]{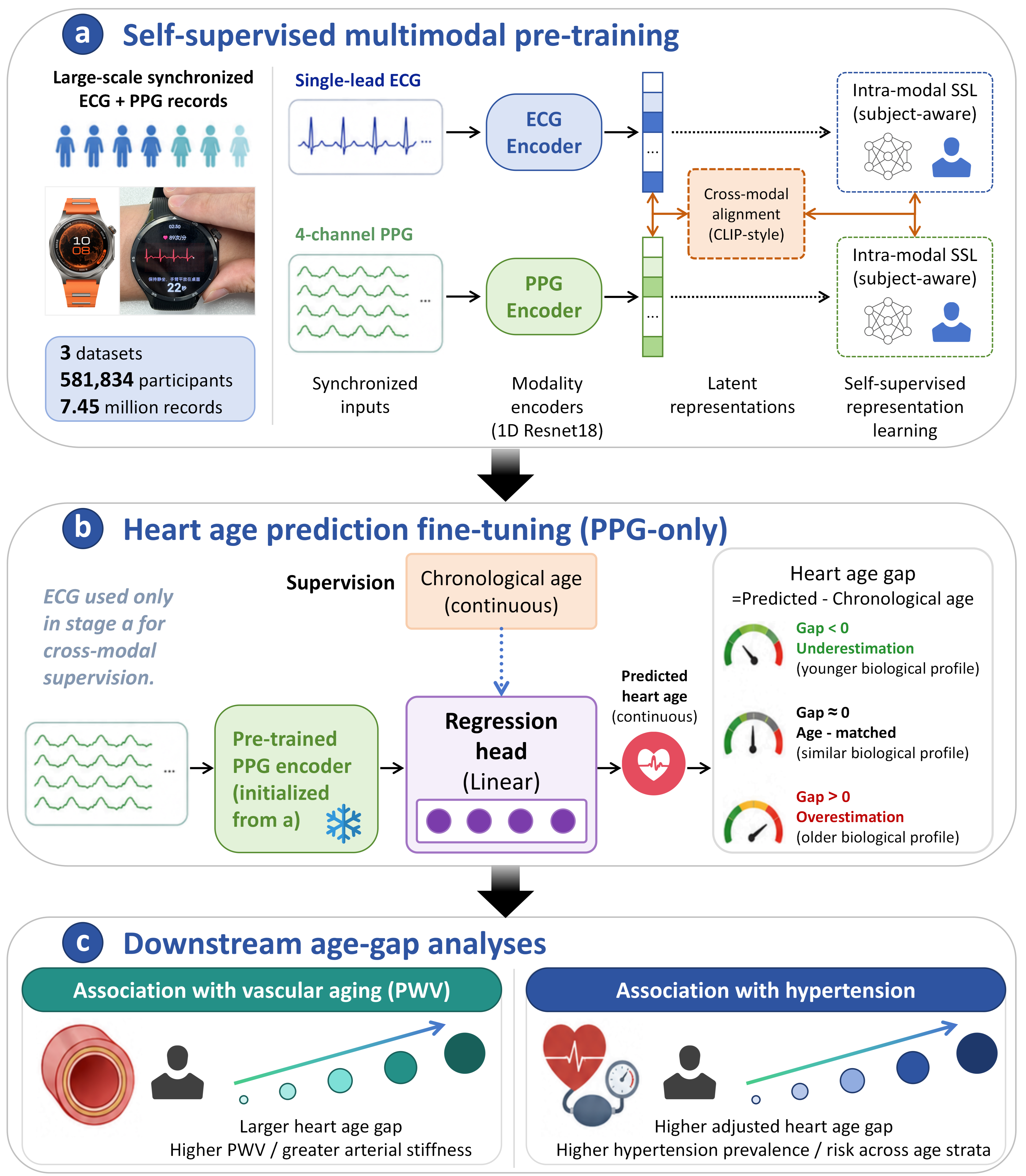}
\caption{
Study overview from ECG-guided cross-modal pretraining to PPG-only heart-age inference and downstream cardiovascular analyses.
\textbf{a}, Self-supervised multimodal pretraining on large-scale synchronized smartwatch ECG--PPG recordings.
Single-lead ECG and 4-channel PPG were encoded by modality-specific 1D ResNet encoders.
The pretraining objective combined subject-aware intra-modal SSL with CLIP-style cross-modal alignment between synchronized ECG--PPG representations.
\textbf{b}, Heart-age prediction fine-tuning using the pretrained PPG encoder with chronological age as supervision.
ECG was used only during pretraining and not required at inference.
The heart age gap (predicted minus chronological age) reflects younger-than-expected, age-matched, or older-than-expected cardiovascular profiles when negative, near-zero, or positive, respectively.
\textbf{c}, The heart age gap was evaluated for associations with vascular aging and arterial stiffness (PWV cohort) and hypertension prevalence and risk (HBPM cohort).
}
\label{fig:study_overview}
\end{figure}

Self-supervised and cross-modal learning offer a methodological route to address these challenges \citep{jaiswal2020survey, krishnan2022self, dunnmon2020cross}. Large-scale unlabeled ECG and PPG signals contain rich individual-specific and cardiovascular temporal information that can be learned through contrastive representation learning. In smartwatch settings, ECG and PPG can be synchronously acquired during active measurements, creating a natural opportunity for ECG-guided cross-modal pretraining. In this study, we propose an ECG-guided cross-modal pretraining framework that uses synchronized ECG and PPG during pretraining to combine subject-aware self-supervised learning with ECG--PPG contrastive alignment. In this design, ECG serves as a privileged modality that helps the PPG encoder learn stronger cardiovascular representations, while inference requires only PPG input to produce smartwatch PPG-derived heart age. We developed and internally evaluated the model in the VHS cohort and then assessed its external generalization and clinical associations in two independent cohorts. An overview of the study design, model training workflow, and downstream age-gap analyses is shown in Fig.~\ref{fig:study_overview}.

The main contributions of this study are as follows.
\begin{enumerate}
    \item We developed a smartwatch PPG-only heart-age estimation framework based on
    synchronized ECG--PPG self-supervised and cross-modal pretraining.
    \item We systematically evaluated model performance across internal and external cohorts
    and quantified the effect of short-term repeated-record aggregation on subject-level
    stability.
    \item We validated the age-independent association between heart age gap and arterial
    stiffness in an external PWV cohort.
    \item We assessed the ability of adjusted heart age gap to stratify hypertension
    prevalence and hypertension odds in an external HBPM cohort.
\end{enumerate}
Through this design, we aimed to establish an end-to-end evidence pathway from wearable
signal representation learning to cardiovascular phenotype validation, and to explore
smartwatch PPG-derived heart age as a scalable digital biomarker of cardiovascular aging.


\section*{RESULTS}


\subsection*{Three-cohort study design enables ECG-guided pretraining, external validation, and cardiovascular association analyses}

This study included three cohorts covering model pretraining, supervised fine-tuning, external validation, and cardiovascular phenotype association analyses. The internal VHS cohort was used for model development. Its unlabeled partition included 536,287 participants and 6,555,251 synchronized ECG--PPG recordings and was used for self-supervised and cross-modal pretraining. Its labeled partition included 43,862 participants and 669,777 recordings and was used for supervised fine-tuning and internal validation of the heart-age prediction task. Two independent external cohorts were used for generalization assessment and clinical association analyses. The external PWV cohort included 560 participants and 4,161 recordings and was used to evaluate the association between model-derived heart age, heart age gap, and pulse wave velocity (PWV). The external HBPM cohort included 1,095 participants and 222,942 recordings and was used to evaluate the association between heart age gap and prevalent hypertension. Across the three cohorts and their analysis partitions, the study included 581,804 participants and 7,452,131 recordings. Detailed information on time frame, signal configuration, sampling rate, and cohort characteristics is provided in Table~\ref{tab:cohort_characteristics}.

The model followed a two-stage design. In the first stage, synchronized ECG and PPG recordings from the VHS cohort were used for pretraining, combining single-modality self-supervised learning with ECG--PPG cross-modal contrastive learning. This design allowed the model to learn both within-modality temporal representations and shared physiological representations across ECG and PPG. In the second stage, the pretrained encoder was fine-tuned on age-labeled VHS data for heart-age regression, using chronological age as the proxy target. Although ECG was used as a cross-modal guiding signal during pretraining, external validation and downstream association analyses primarily used the PPG-only inference pathway, reflecting the intended smartwatch deployment scenario. Based on the model output, we defined predicted heart age and further calculated heart age gap as the difference between predicted heart age and chronological age. We then examined the association between heart age gap and arterial stiffness in the PWV cohort and evaluated its ability to stratify hypertension prevalence and odds in the HBPM cohort.


\begin{table}[htbp]
\centering
\caption{%
Study cohorts, signal specifications, and participant characteristics.
VHS denotes vascular health study; PWV, pulse wave velocity;
HBPM, home blood pressure monitoring; ECG, electrocardiography;
PPG, photoplethysmography; BMI, body mass index.
The VHS cohort was used for model development and was divided into
an unlabeled pretraining partition and a labeled fine-tuning/internal
validation partition.
The PWV cohort was used for external validation and arterial stiffness
association analyses.
The HBPM cohort was used for external validation and hypertension
association analyses.
Values are reported as mean $\pm$ standard deviation or number (percentage).
N/A indicates variables that were not used or not available in the
corresponding cohort or analysis partition.}
\label{tab:cohort_characteristics}
\footnotesize
\setlength{\tabcolsep}{5pt}
\renewcommand{\arraystretch}{1.3}
\newcolumntype{C}{>{\centering\arraybackslash}p{3.2cm}}
\newcolumntype{L}{>{\raggedright\arraybackslash}p{3.6cm}}
\begin{tabular}{LCCCC}
\toprule
\textbf{Characteristic}
  & \textbf{VHS Unlabeled}
  & \textbf{VHS Labeled}
  & \textbf{PWV (External)}
  & \textbf{HBPM (External)} \\
\midrule

Time frame
  & 2022--2025
  & 2022--2025
  & 2023--2025
  & 2022--2024 \\[5pt]

Signal channels / sampling rate
  & 1-ch ECG + 4-ch PPG (green); 250~Hz
  & 1-ch ECG + 4-ch PPG (green); 250~Hz
  & 1-ch ECG + 4-ch PPG (green); 100/250~Hz
  & 2- or 4-ch PPG (green); 250~Hz \\[5pt]

Participants / records
  & 536,287 / 6,555,251
  & 43,862 / 669,777
  & 560 / 4,161
  & 1,095 / 222,942 \\[5pt]

Female, $n$ (\%)
  & N/A
  & 4,912 (11.2\%)
  & 211 (37.7\%)
  & 546 (49.9\%) \\[5pt]

Age, years
  & N/A
  & $32.37 \pm 9.20$
  & $40.68 \pm 14.71$
  & $37.73 \pm 9.52$ \\[5pt]

BMI, kg/m$^{2}$
  & N/A
  & $24.89 \pm 4.23$
  & $24.36 \pm 3.26$
  & $24.09 \pm 3.64$ \\[5pt]

PWV, m/s
  & N/A
  & N/A
  & $8.41 \pm 1.85$
  & N/A \\[5pt]

Hypertension, $n$ (\%)
  & N/A
  & N/A
  & N/A
  & Total: 342 (31.2\%) \newline
    F: 159 (29.1\%) \newline
    M: 183 (33.3\%) \\

\bottomrule
\end{tabular}
\end{table}

\subsection*{ECG-guided cross-modal pretraining improves subject-level PPG heart-age estimation and generalizes to external cohorts}

In the internal VHS cohort, we first compared single-modality self-supervised pretraining (SA-SSL) with ECG--PPG cross-modal contrastive pretraining (SA-CLIP) for heart-age estimation across different input modalities. Model performance was evaluated at both the record and subject levels, with subject-level estimates derived by aggregating repeated recordings from the same participant. Compared with 1-lead ECG, 4-channel PPG achieved better performance after SA-SSL pretraining, with a subject-level MAE of 5.223 years, Pearson's \textit{r} of 0.687, and $R^2$ of 0.453; the corresponding values for the 1-lead ECG model were 5.884 years, 0.553, and 0.299, respectively.

After introducing SA-CLIP, the PPG-only model showed lower subject-level error and higher association with chronological age. Compared with PPG SA-SSL, PPG SA-CLIP reduced the subject-level MAE from 5.223 to 4.951 years and the RMSE from 6.792 to 6.394 years, while increasing Pearson's \textit{r} from 0.687 to 0.734 and $R^2$ from 0.453 to 0.521. At the record level, PPG SA-CLIP also achieved lower MAE and RMSE than PPG SA-SSL, with values of 6.084 and 8.074 years compared with 6.141 and 8.151 years, respectively. A similar subject-level improvement was observed for the ECG branch, where the MAE decreased from 5.884 to 5.574 years and Pearson's \textit{r} increased from 0.553 to 0.637.

The multimodal fusion model achieved a subject-level MAE of 4.910 years and Pearson's \textit{r} of 0.730 in the internal VHS cohort. However, its subject-level correlation and $R^2$ did not exceed those of the PPG-only SA-CLIP model. Given the overall subject-level performance of PPG SA-CLIP and its compatibility with smartwatch-only deployment without ECG input at inference, this 4-channel PPG-only model was selected for subsequent external validation.

In two independent external cohorts, the PPG SA-CLIP model maintained stable predictive performance. In the external PWV cohort, the model achieved a subject-level MAE of 5.895 years, Pearson's \textit{r} of 0.819, and $R^2$ of 0.571. In the external HBPM cohort, the corresponding values were 4.344 years, 0.800, and 0.588. Compared with record-level estimates, subject-level aggregation reduced MAE and RMSE and improved Pearson's \textit{r} and $R^2$ across all cohorts. Taken together, these results indicate that ECG-guided cross-modal pretraining strengthens PPG representations for subject-level heart-age estimation and that the PPG-only model generalizes robustly to independent PWV and HBPM cohorts.


\begin{table}
\centering
\caption{
Performance of heart-age estimation in the internal VHS and external PWV/HBPM cohorts. Performance is reported as record-level / subject-level metrics. MAE and RMSE are reported in years.
VHS denotes Vascular Health Study; PWV, pulse wave velocity; HBPM, home blood pressure monitoring.
SA-SSL denotes single-modality self-supervised pretraining, and SA-CLIP denotes ECG--PPG cross-modal contrastive pretraining.
The internal multimodal fusion model is included as a benchmark.
External validation in the PWV and HBPM cohorts was conducted using the SA-CLIP-pretrained 4-channel PPG-only model; ECG was used only during cross-modal pretraining and was not required for external inference.
}
\label{tab:heart_age_performance}
\scriptsize
\setlength{\tabcolsep}{4pt}
\renewcommand{\arraystretch}{1.18}
\resizebox{\textwidth}{!}{
\begin{tabular}{lllcccc}
\toprule
\textbf{Cohort} &
\textbf{Modality} &
\textbf{Pretraining} &
\makecell{\textbf{MAE} $\downarrow$\\\textbf{(record / subject)}} &
\makecell{\textbf{RMSE} $\downarrow$\\\textbf{(record / subject)}} &
\makecell{\textbf{Pearson's \textit{r}} $\uparrow$\\\textbf{(record / subject)}} &
\makecell{\textbf{$R^2$} $\uparrow$\\\textbf{(record / subject)}} \\
\midrule
\multirow{5}{*}{\makecell{Internal\\VHS}}
& 1ECG & SA-SSL
& 6.736 / 5.884
& 8.911 / 7.688
& 0.525 / 0.553
& 0.268 / 0.299 \\
& 4PPG & SA-SSL
& 6.141 / 5.223
& 8.151 / 6.792
& 0.623 / 0.687
& 0.388 / 0.453 \\
& 1ECG & SA-CLIP
& 6.516 / 5.574
& 8.470 / 7.186
& 0.507 / 0.637
& 0.237 / 0.395 \\
& 4PPG & SA-CLIP
& 6.084 / 4.951
& 8.074 / 6.394
& 0.577 / 0.734
& 0.306 / 0.521 \\
& 1ECG + 4PPG & SA-CLIP
& 5.941 / 4.910
& 7.889 / 6.411
& 0.592 / 0.730
& 0.338 / 0.518 \\
\midrule
\makecell{External\\PWV}
& 4PPG & SA-CLIP
& 6.927 / 5.895
& 11.038 / 9.628
& 0.788 / 0.819
& 0.505 / 0.571 \\
\makecell{External\\HBPM}
& 4PPG & SA-CLIP
& 5.247 / 4.344
& 7.187 / 6.103
& 0.703 / 0.800
& 0.448 / 0.588 \\
\bottomrule
\end{tabular}
}
\vspace{2mm}
\begin{minipage}{0.98\textwidth}
\footnotesize
\end{minipage}
\end{table}


\begin{figure}
\centering
\includegraphics[width=\textwidth]{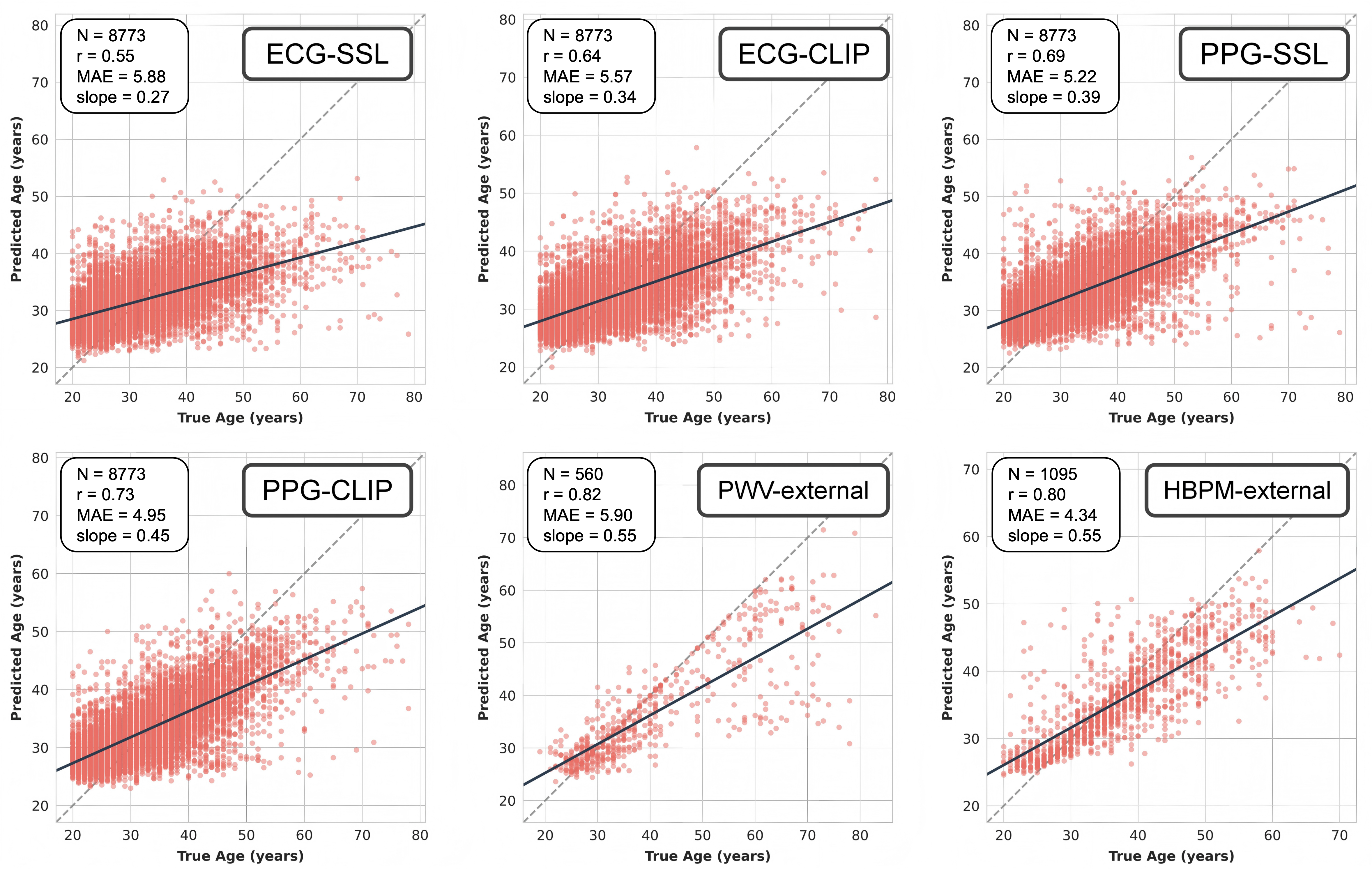}
\caption{
Heart-age estimation performance across internal and external cohorts.
Scatter plots show subject-level predicted heart age versus chronological age.
The internal VHS cohort includes 1-lead ECG and 4-channel PPG models pretrained with SA-SSL or SA-CLIP; the PWV-external and HBPM-external panels show external validation results.
External validation in the PWV and HBPM cohorts was performed using the SA-CLIP-pretrained 4-channel PPG-only model; ECG was used only during cross-modal pretraining and was not used at inference.
Each point represents one participant after subject-level aggregation.
The dashed line indicates the identity line, and the solid line indicates the fitted linear trend.
Pearson's \textit{r}, MAE, and fitted slope are reported within each panel.
}
\label{fig:heart_age_performance}
\end{figure}

\subsection*{Short-term multi-record aggregation improves and stabilizes subject-level
heart-age estimation}

In the internal VHS cohort, we further evaluated the effect of short-term multi-record
aggregation on subject-level heart-age estimation. Because the repeated recordings included
in this analysis were acquired within a short-term window of no more than 3 months, we
assumed that each participant remained in a relatively stable physiological state during
this period and that the underlying heart-age phenotype was therefore approximately stable.
We then averaged predictions from multiple recordings within each participant and evaluated
subject-level performance as the number of included recordings increased from 1 to 15.
As the number of aggregated recordings increased, prediction error decreased, while
correlation and explained variance improved. Specifically, when only one recording per
participant was used, the MAE was 5.395 years, the RMSE was 7.069 years, Pearson's
\textit{r} was 0.649, and $R^2$ was 0.414. When 15 recordings were aggregated, the MAE
decreased to 4.949 years and the RMSE decreased to 6.393 years, while Pearson's \textit{r}
increased to 0.733 and $R^2$ increased to 0.521. The largest improvement occurred during
the aggregation of the first few recordings, followed by a gradual plateau, suggesting that
short-term repeated measurements can reduce single-record variability and yield a more
stable subject-level estimate of heart age (see Supplementary Table~\ref{tab:scaling_law_multisample}).

We next performed a data-density threshold analysis by restricting the cohort to
participants with at least $N_{\min}$ recordings and then calculating subject-level
performance after aggregation. As $N_{\min}$ increased from 1 to 15, the number of
included participants decreased from 8773 to 2447. Prediction error reached a lower range
at moderate recording densities and remained relatively stable at higher thresholds,
whereas Pearson's \textit{r} and $R^2$ continued to increase with greater data density.
At a threshold of at least 10 recordings per participant, 3463 participants remained, and
the model achieved an MAE of 4.930 years, an RMSE of 6.198 years, Pearson's \textit{r}
of 0.806, and $R^2$ of 0.582. When the threshold was further increased to at least 15
recordings, Pearson's \textit{r} and $R^2$ increased to 0.817 and 0.592, respectively
(see Supplementary Table~\ref{tab:threshold_density}). Together, these two analyses indicate that, under the
assumption of short-term physiological stability, aggregating repeated wearable recordings
improves the accuracy and stability of subject-level heart-age estimation and supports
repeated-measurement reporting in smartwatch-based applications.


\begin{figure}
\centering
\includegraphics[width=\textwidth]{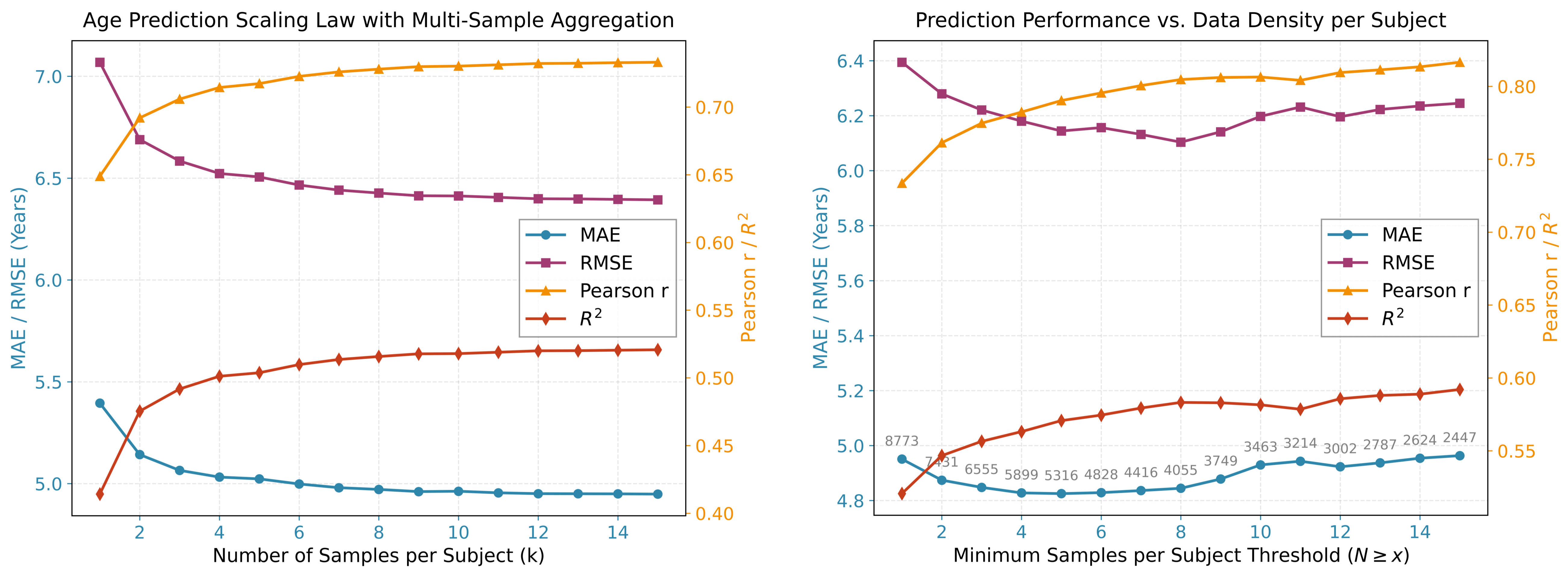}
\caption{
Effect of short-term multi-record aggregation on heart-age estimation performance.
The left panel shows changes in subject-level MAE, RMSE, Pearson's \textit{r}, and $R^2$ as the number of aggregated recordings per participant, $k$, increases from 1 to 15.
The right panel shows model performance after retaining only participants with at least $N_{\min}$ recordings, together with the number of remaining participants.
All repeated recordings were acquired within a short-term window of no more than 3 months; the analysis assumes that each participant's heart-age phenotype is approximately stable over this interval.
Lower MAE and RMSE indicate lower prediction error, whereas higher Pearson's \textit{r} and $R^2$ indicate stronger agreement between predicted and chronological age.
}
\label{fig:scaling_law}
\end{figure}


\subsection*{Heart age gap is independently associated with arterial stiffness beyond chronological age}

In the external PWV cohort, we further evaluated whether the model-derived heart age gap captured arterial stiffness information beyond chronological age. Heart age gap was defined as the difference between predicted heart age and chronological age, 
$\Delta_{\mathrm{HA}}=\widehat{\mathrm{Age}}_{\mathrm{heart}}-\mathrm{Age}_{\mathrm{chron}}$.
A positive gap indicates that the model predicted an older heart age than the participant's chronological age, whereas a negative gap indicates a younger predicted heart age.

\begin{figure}
\centering
\includegraphics[width=0.6\textwidth]{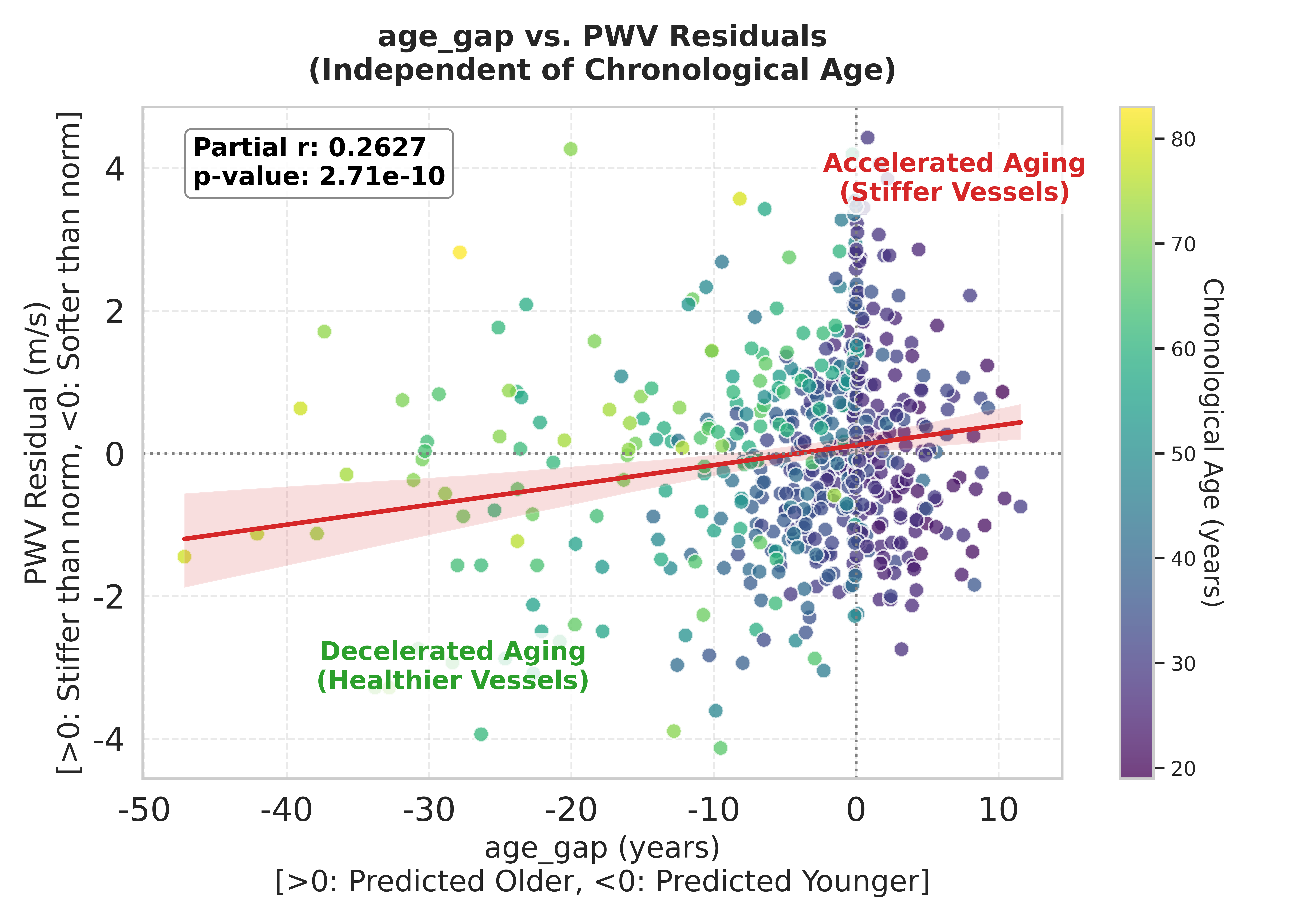}
\caption{
Age-independent association between heart age gap and PWV residuals.
The scatter plot shows the association between heart age gap and PWV residuals.
PWV residuals were obtained after removing the effect of chronological age on PWV; point color indicates chronological age.
Dashed lines indicate zero residual and zero heart age gap.
The red line represents the fitted linear trend, with the shaded area indicating the 95\% confidence interval.
The reported partial correlation quantifies the association between heart age gap and PWV after adjustment for chronological age.
}
\label{fig:pwv_partial_corr}
\end{figure}

We first used partial correlation analysis to examine the association between $\Delta_{\mathrm{HA}}$ and PWV while adjusting for chronological age. After accounting for chronological age, heart age gap remained significantly and positively associated with PWV, with a partial correlation coefficient of 0.2627 and $P=2.71\times10^{-10}$ (Fig.~\ref{fig:pwv_partial_corr}). In the residual-based visualization, positive heart age gaps corresponded to higher PWV residuals, indicating greater arterial stiffness than expected for chronological age, whereas negative gaps corresponded to lower PWV residuals.

To further quantify this association, we fitted a multivariable linear regression model with PWV as the continuous outcome and chronological age, sex, and BMI as covariates (Table~\ref{tab:pwv_ols}). After adjustment, each 1-year increase in heart age gap was associated with a 0.062 m/s higher PWV (95\% CI: 0.040--0.084, $P<0.001$). Accordingly, among participants with the same chronological age, sex, and BMI, a 5-year higher heart age gap corresponded to an approximately 0.31 m/s higher PWV. The model explained 52.5\% of the variance in PWV, with an adjusted $R^2$ of 0.521. These results indicate that heart age gap provides arterial stiffness information independent of chronological age and conventional anthropometric factors.


\begin{table}[htbp]
\centering
\caption{Multivariable linear regression analysis of PWV.}
\label{tab:pwv_ols}
\small
\setlength{\tabcolsep}{18pt} 
\renewcommand{\arraystretch}{1.3} 
\begin{threeparttable}
\begin{tabular}{lccc}
\toprule
\textbf{Variable} & \textbf{$\beta$} & \textbf{95\% CI} & \textbf{\textit{P} value} \\
\midrule
Heart age gap, years \tnote{a} & 0.062 & 0.040--0.084 & $<0.001$ \\
Chronological age, years       & 0.106 & 0.094--0.118 & $<0.001$ \\
Sex, male vs. female           & 0.853 & 0.611--1.095 & $<0.001$ \\
BMI, kg/m$^2$                  & 0.062 & 0.027--0.098 & $<0.001$ \\
\bottomrule
\end{tabular}
\vspace{2mm}
\begin{tablenotes}[flushleft]
\footnotesize
\setlength{\itemsep}{1pt} 
\item \textit{Note.} The outcome variable was PWV in m/s. $\beta$ denotes the unstandardized regression coefficient. Confidence intervals are reported as 95\% CIs. Heteroscedasticity-robust standard errors were used. Model statistics: $N=560$, $R^2=0.525$, adjusted $R^2=0.521$, $F(4,555)=167.4$, $P<0.001$.
\item[a] Heart age gap was defined as predicted heart age minus chronological age.
\end{tablenotes}
\end{threeparttable}
\end{table}


\subsection*{Accelerated heart aging stratifies individuals by adjusted arterial stiffness}

\begin{figure}
\centering
\includegraphics[width=0.55\textwidth]{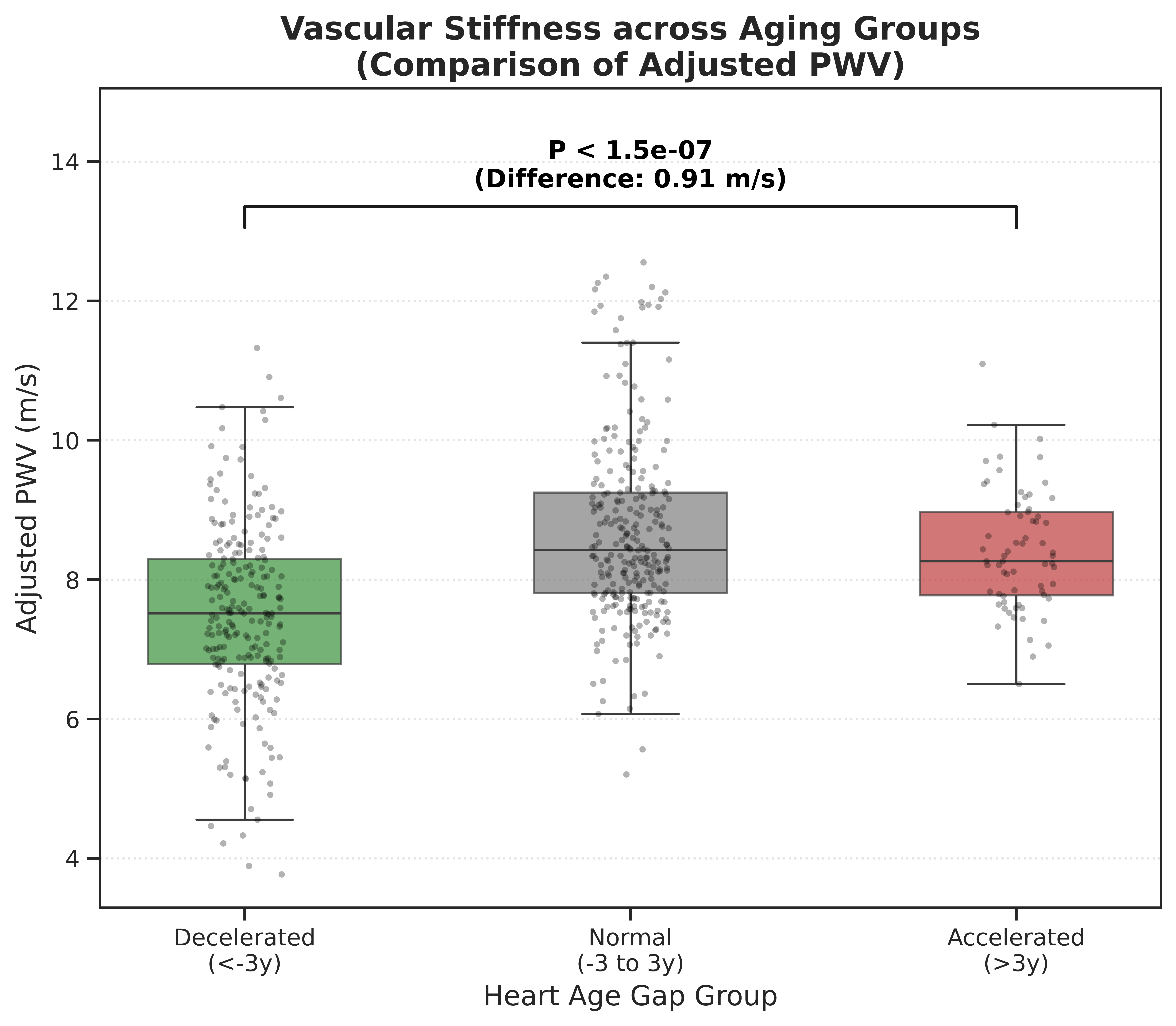}
\caption{
\textbf{Distribution of adjusted PWV across heart age gap groups.}
Participants were categorized into decelerated heart aging ($<-3$ years), normal aging ($-3$ to $3$ years), and accelerated heart aging ($>3$ years) according to heart age gap.
Box plots show the distribution of adjusted PWV across groups, with overlaid points representing individual participants.
Adjusted PWV denotes PWV after adjustment for chronological age, sex, and BMI.
The bracket denotes the comparison between the decelerated and accelerated groups.
The accelerated group had a 0.91 m/s higher adjusted PWV than the decelerated group ($P<0.001$).
}
\label{fig:pwv_groups}
\end{figure}

To improve clinical interpretability, we categorized participants into three groups according to heart age gap: decelerated heart aging ($<-3$ years), normal aging ($-3$ to $3$ years), and accelerated heart aging ($>3$ years). We then compared the distribution of adjusted PWV across these groups (Fig.~\ref{fig:pwv_groups}). Adjusted PWV increased across the decelerated, normal, and accelerated heart-aging groups. Participants in the accelerated heart-aging group had significantly higher adjusted PWV than those in the decelerated heart-aging group, with a between-group difference of 0.91 m/s.

As descriptive cohort context, 136 of the 560 participants in the external PWV cohort exceeded the prespecified PWV threshold of $>10.0$ m/s, corresponding to 24.3\% of the cohort. This indicates a substantial burden of elevated arterial stiffness in the external PWV cohort; however, the primary evidence in this section is based on continuous PWV analyses, including partial correlation, multivariable regression, and group-wise comparison of adjusted PWV. Overall, heart age gap was not only independently associated with PWV in statistical models but also provided a simple and interpretable stratification of vascular stiffness phenotypes.


\subsection*{Adjusted heart age gap differentiates hypertension prevalence across age strata}

In the external HBPM cohort, we further evaluated whether the heart age gap could stratify hypertension prevalence within chronological-age groups. To reduce systematic age-related bias in the raw heart age gap, we corrected the heart age gap using an age-based generalized additive model with a B-spline smoothing term. Subsequent analyses were based on the adjusted heart age gap, denoted as $\Delta_{\mathrm{HA,adj}}$. Participants were stratified by chronological age into $<30$, $[30,40)$, $[40,50)$, and $[50,60)$ years and further grouped by $\Delta_{\mathrm{HA,adj}}$ into $\leq0$, $(0,2]$, $(2,4]$, and $>4$ years. For each age-by-gap stratum, we estimated hypertension prevalence, bootstrap 95\% confidence intervals, and relative prevalence compared with the average prevalence within the same chronological-age group (Fig.~\ref{fig:hypertension_gap_strata}; detailed values are provided in Supplementary Table~\ref{tab:supp_hypertension_gap}).

Chronological age remained the dominant background determinant of hypertension prevalence. Across the four age strata shown in Fig.~\ref{fig:hypertension_gap_strata}, the average prevalence of hypertension increased with age, from 12.8\% in participants aged $<30$ years to 25.4\% in those aged $[30,40)$ years, 42.9\% in those aged $[40,50)$ years, and 58.3\% in those aged $[50,60)$ years. Within these chronological-age strata, adjusted heart age gap provided additional prevalence stratification. Among all participants aged $<30$ years, hypertension prevalence increased from 7.2\% in the $\Delta_{\mathrm{HA,adj}}\leq0$ group to 48.0\% in the $>4$-year group, corresponding to a relative prevalence of 3.74. In the $[30,40)$-year group, prevalence was 14.7\% in the $\leq0$ group, compared with 54.3\% and 43.5\% in the $(2,4]$-year and $>4$-year groups, respectively. In the $[40,50)$-year group, prevalence increased from 32.5\% in the $\leq0$ group to 60.0\% in the $>4$-year group. In the $[50,60)$-year group, the two higher-gap groups had prevalences of 78.3\% and 68.2\%, both higher than the 50.9\% observed in the $\leq0$ group.

Sex-stratified analyses showed a similar directional pattern. Among women, the gradient between adjusted heart age gap and hypertension prevalence was more continuous. In women aged $<30$ years, prevalence increased from 4.3\% in the $\leq0$ group to 46.7\% in the $>4$-year group, corresponding to a relative prevalence of 3.58. In women aged $[30,40)$ years, prevalence was 12.4\% in the $\leq0$ group and 47.8\% and 46.7\% in the $(2,4]$-year and $>4$-year groups, respectively. In women aged $[40,50)$ years, prevalence increased from 14.7\% in the $\leq0$ group to 61.8\% in the $>4$-year group. Among men, higher-gap strata also showed elevated prevalence. For example, in men aged $<30$ years, the $>4$-year group had a prevalence of 50.0\% and a relative prevalence of 3.97; in men aged $[30,40)$ years, the $(2,4]$-year group had a prevalence of 66.7\% and a relative prevalence of 2.67. Some male high-gap strata had small sample sizes, resulting in wider confidence intervals.

Lower or negative adjusted heart age gap was associated with lower hypertension prevalence in most age strata. Among all participants, the $\Delta_{\mathrm{HA,adj}}\leq0$ group had relative prevalences of 0.56, 0.58, 0.76, and 0.87 across the four chronological-age strata, all below or close to the age-group average. This low-prevalence pattern was particularly evident among women, where the corresponding relative prevalences were 0.33, 0.48, 0.42, and 0.77. Taken together, these stratified analyses indicate that adjusted heart age gap further differentiates hypertension prevalence beyond chronological age, with a clear gradient particularly among younger and middle-aged participants.


\begin{figure}[htbp]
\centering
\includegraphics[
    width=\textwidth,
    height=0.75\textheight,
    keepaspectratio
]{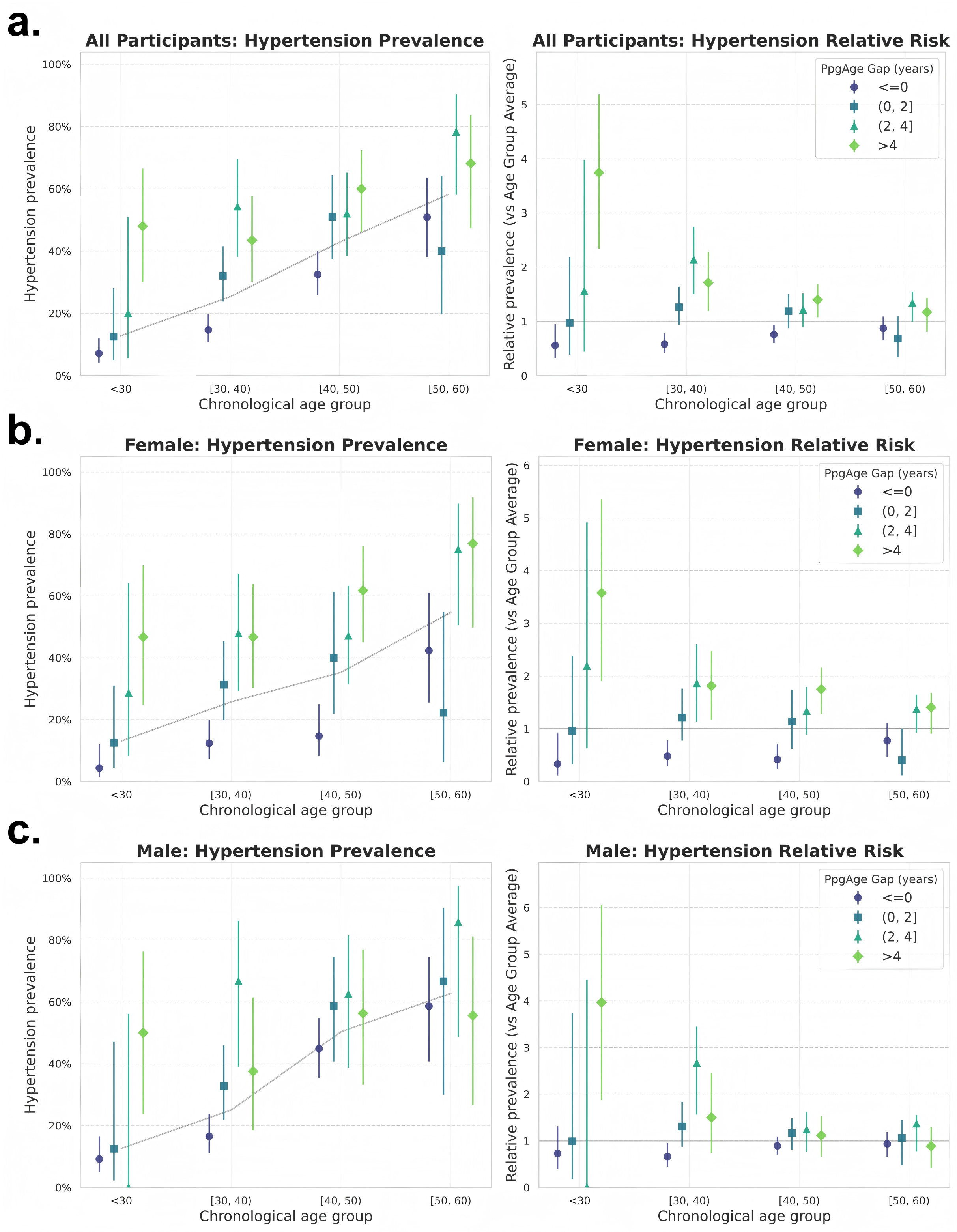}
\caption{
Hypertension prevalence across chronological-age and adjusted heart age gap strata in the HBPM cohort.
Participants were stratified by chronological age into $<30$, $[30,40)$, $[40,50)$, and $[50,60)$ years and by adjusted heart age gap $\Delta_{\mathrm{HA,adj}}$ into $\leq0$, $(0,2]$, $(2,4]$, and $>4$ years.
The left column shows hypertension prevalence within each stratum, with error bars indicating bootstrap 95\% confidence intervals; the gray line indicates the average prevalence within each chronological-age group.
The right column shows relative prevalence, defined as the prevalence within each gap stratum divided by the average prevalence within the same chronological-age group; the horizontal line indicates a relative prevalence of 1.
Panels a, b, and c show all participants, women, and men, respectively.
Because the HBPM cohort is cross-sectional, the estimates represent prevalence and relative prevalence rather than prospective incidence risk.
}
\label{fig:hypertension_gap_strata}
\end{figure}


\subsection*{Adjusted heart age gap remains independently associated with hypertension odds after covariate adjustment}

To further test whether the prevalence stratification observed in the previous section persisted after adjustment for conventional clinical covariates, we performed multivariable logistic regression analyses in the HBPM cohort. The analytic sample was restricted to participants with complete data on hypertension status, adjusted heart age gap, chronological age, BMI, sex, and family history of hypertension, resulting in 1095 participants. Hypertension status was used as the binary outcome, and the adjusted heart age gap, denoted as $\Delta_{\mathrm{HA,adj}}$, was the primary explanatory variable. The overall model was adjusted for chronological age, sex, BMI, and family history of hypertension, whereas sex-stratified models were adjusted for chronological age, BMI, and family history of hypertension. We modeled $\Delta_{\mathrm{HA,adj}}$ both as a standardized continuous variable and as quartiles (Fig.~\ref{fig:hypertension_or}; detailed values are provided in Supplementary Table~\ref{tab:supp_hypertension_quartile}).

In the continuous model, each 1-SD increase in $\Delta_{\mathrm{HA,adj}}$ was significantly associated with higher odds of prevalent hypertension. Among all participants, the OR per 1-SD increase was 1.72 (95\% CI: 1.49--1.99, $P<0.001$). Sex-stratified analyses showed that this association was present in both men and women, with ORs of 1.52 (95\% CI: 1.24--1.87, $P<0.001$) and 1.96 (95\% CI: 1.59--2.43, $P<0.001$), respectively. These findings indicate that, when treated as a continuous digital phenotype, higher adjusted heart age gap is independently associated with higher odds of prevalent hypertension.

The quartile-based model further showed that the increase in hypertension odds was concentrated in the higher heart age gap quartiles. Using the lowest quartile, Q1, as the reference group, Q2 was not significantly different from Q1 in the overall cohort (OR: 1.01, 95\% CI: 0.63--1.61, $P=0.96$), whereas Q3 and Q4 were associated with significantly higher odds of hypertension, with ORs of 2.38 (95\% CI: 1.56--3.62, $P<0.001$) and 4.25 (95\% CI: 2.80--6.46, $P<0.001$), respectively. The observed prevalence of hypertension increased from 21\% in Q1 (57/274) to 54\% in Q4 (147/274). Among men, the ORs for Q3 and Q4 relative to Q1 were 2.55 (95\% CI: 1.48--4.40, $P<0.001$) and 3.03 (95\% CI: 1.69--5.45, $P<0.001$), respectively. Among women, the corresponding ORs were 2.43 (95\% CI: 1.19--4.94, $P=0.014$) and 5.63 (95\% CI: 2.91--10.88, $P<0.001$). The point estimate for the highest quartile was larger in women than in men, although this sex difference should be interpreted descriptively unless formally tested using an interaction term.

Taken together, the logistic regression analyses complement the stratified prevalence results from the previous section: the former quantifies the covariate-adjusted strength of association, whereas the latter illustrates absolute prevalence differences within chronological-age strata. Because the HBPM cohort is cross-sectional, the ORs should be interpreted as odds ratios for prevalent hypertension rather than prospective incidence risk.


\begin{figure}[htbp]
\centering
\includegraphics[
    width=\textwidth,
    height=0.75\textheight,
    keepaspectratio
]{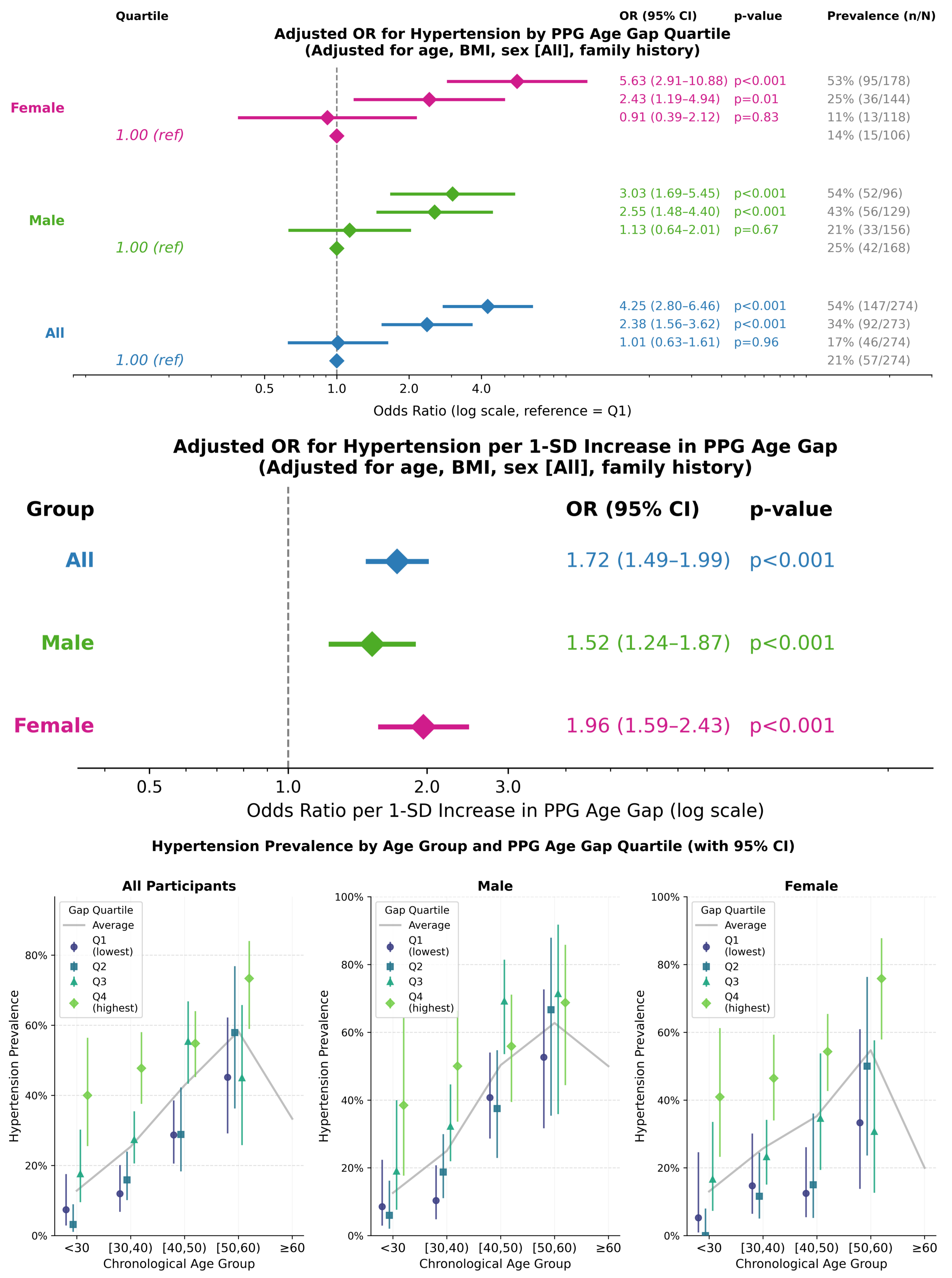}
\caption{
Multivariable logistic regression analysis of adjusted heart age gap and hypertension odds.
\textbf{a}, Adjusted odds ratios for hypertension across quartiles of adjusted heart age gap, $\Delta_{\mathrm{HA,adj}}$.
The lowest quartile, Q1, was used as the reference group.
The overall model was adjusted for chronological age, sex, BMI, and family history of hypertension; sex-stratified models were adjusted for chronological age, BMI, and family history of hypertension.
The x-axis is shown on a log scale, and the dashed vertical line indicates OR = 1.
The right columns show ORs, 95\% CIs, $P$ values, and the observed prevalence in each quartile.
\textbf{b}, Adjusted odds ratios for hypertension per 1-SD increase in $\Delta_{\mathrm{HA,adj}}$.
\textbf{c}, Descriptive hypertension prevalence across chronological-age groups and $\Delta_{\mathrm{HA,adj}}$ quartiles, shown to contextualize the logistic regression results.
Error bars indicate 95\% CIs, and gray lines indicate the average prevalence within each chronological-age group.
Because the HBPM cohort is cross-sectional, the estimates represent odds of prevalent hypertension and observed prevalence rather than prospective incidence risk.
}
\label{fig:hypertension_or}
\end{figure}


\section*{DISCUSSION}

In this study, we developed and validated a smartwatch PPG-based heart-age estimation framework that uses ECG-guided cross-modal pretraining to enhance PPG representation learning while ultimately supporting PPG-only inference. Using large-scale physiological signal data from three cohorts, we first performed model pretraining, supervised fine-tuning, and internal evaluation in the VHS cohort, followed by external validation in two independent cohorts and downstream association analyses with arterial stiffness and prevalent hypertension. The main findings were fourfold. First, ECG--PPG cross-modal pretraining improved subject-level PPG-only heart-age estimation and maintained stable performance in the external PWV and HBPM cohorts. Second, short-term aggregation of repeated recordings further improved the accuracy and stability of subject-level estimates. Third, heart age gap remained independently associated with PWV after adjustment for chronological age, sex, and BMI, and stratified individuals by adjusted arterial stiffness. Fourth, adjusted heart age gap was consistently associated with both hypertension prevalence and hypertension odds. Taken together, these findings establish an end-to-end evidence from wearable signal representation learning and PPG-only heart-age inference to external cardiovascular phenotype validation, supporting PPG-derived heart age as a scalable digital phenotype of cardiovascular aging.

From an algorithmic perspective, our results suggest that smartwatch PPG may capture more age-related peripheral hemodynamic information than single-lead smartwatch ECG for the heart-age estimation task. In the internal VHS cohort, the PPG-only model generally outperformed the ECG-only model, indicating that pulse morphology, peripheral vascular compliance, and hemodynamic changes encoded in PPG may be sensitive to age-related phenotypes \cite{charlton2022assessing, fathieh2021predicting}. More importantly, ECG-guided cross-modal pretraining further improved the subject-level performance of the PPG-only model. This supports a training paradigm in which multimodal information is used during pretraining, whereas a single modality is retained for deployment. During pretraining, synchronized ECG serves as an additional physiological supervisory signal that helps the PPG encoder learn representations related to cardiovascular rhythm, beat-to-beat dynamics, and shared cross-modal physiological states. During external validation and practical use, however, the model requires only PPG input for heart-age estimation. In this sense, ECG acts as a privileged modality or a privileged supervisory signal that transfers information to the PPG representation during training without increasing the sensor burden at inference \cite{wei2025ecg}. This design is particularly relevant for smartwatch applications, because PPG can usually be collected passively, repeatedly, and with low user burden, whereas ECG often requires active measurement or a specific contact posture. ECG-guided cross-modal pretraining therefore provides a practical route to improving both the representation quality and deployability of wearable PPG models.

This work also extends the use of heart-age and vascular-age concepts into the wearable device setting. Existing heart-age or vascular-age frameworks have often relied on conventional risk factors, ECG, imaging, or clinical vascular function measurements, with the shared goal of translating complex cardiovascular status into an interpretable age scale \cite{ladejobi202112, park2024artificial, cho2025artificial, hempel2025explainable, al2025advanced, raghu2021deep}. Our findings show that heart age gap inferred from smartwatch PPG alone can provide vascular aging information beyond chronological age. In the external PWV cohort, heart age gap was significantly associated with PWV residuals and remained independently associated with PWV in multivariable OLS regression. In group-wise analyses, individuals with accelerated heart aging had higher adjusted PWV. These results suggest that PPG-derived heart age is not merely a reconstruction of chronological age, but captures residual information related to arterial stiffness. Clinically, such a residual aging signal may be useful: among individuals of the same chronological age, a higher PPG-derived heart age may indicate earlier or more pronounced vascular aging. Importantly, this measure should not be interpreted as a directly measured ``true biological age''. Rather, heart age gap represents a model-derived deviation from expected age-related signal patterns learned from PPG.

The association between PPG-derived heart age and vascular aging is physiologically plausible. PPG records optical pulse volume changes in the peripheral microvascular bed across the cardiac cycle, and its waveform is influenced by arterial compliance, peripheral resistance, pulse wave propagation, reflected wave timing, vascular tone, and autonomic regulation. With aging and increasing arterial stiffness, pulse waves propagate faster, reflected waves return earlier, and peripheral pulse morphology may change in systolic upstroke, peak shape, diastolic decay, and reflection-related components \cite{yousef2012analysis, huotari2011photoplethysmography}. Thus, although PPG is not a direct measurement of PWV, its waveform may encode integrated information about vascular elasticity and hemodynamic state. The observed associations of heart age gap with PWV and hypertension support this interpretation: the model may learn age-related vascular and hemodynamic signatures from PPG and transform them into a digital phenotype useful for cardiovascular risk stratification. In the HBPM cohort, higher adjusted heart age gap corresponded to higher hypertension prevalence within multiple chronological-age strata and remained independently associated with hypertension odds after adjustment for chronological age, BMI, sex, and family history. This phenotype may be particularly relevant in younger and middle-aged adults, for whom conventional risk assessment is often dominated by chronological age, while heart age gap may help identify individuals with already elevated vascular or blood pressure burden despite relatively young chronological age \cite{luo2020association, laurent2019concept}.

The scaling-law analysis of short-term repeated recordings further highlights the practical value of moving from single-time-point testing to continuous monitoring in wearable applications. A single PPG recording can be affected by device wearing condition, motion artifact, skin contact, ambient temperature, and transient physiological fluctuations, and therefore inevitably contains random prediction error \cite{pollreisz2022detection}. By aggregating repeated recordings from the same participant within a window of no more than 3 months, we observed progressive reductions in subject-level MAE and RMSE and corresponding increases in Pearson's \(r\) and \(R^2\). Performance became relatively stable when approximately 10 or more recordings were available per participant. This finding has direct implications for product design: smartwatch-based heart age should not necessarily be reported from a single PPG measurement, but can instead be estimated after accumulating multiple high-quality recordings within a short-term stable window and aggregating them at the individual level \cite{viswanath2026sleep}. In future deployment, the model output could be reported together with the number of qualified recordings, signal quality, and estimation uncertainty, improving both user interpretability and clinical trustworthiness.

Several limitations should be acknowledged. First, the three cohorts were derived from specific devices and populations, with a predominance of Asian and relatively young participants and a higher proportion of men in the internal development cohort; therefore, generalizability to older adults, multi-ethnic populations, individuals with more comorbidities, and different device settings requires further validation. Second, the heart-age model was trained using chronological age as a proxy target rather than a directly measured biological-age ground truth \cite{salih2023conceptual}. Accordingly, heart age gap should be interpreted as a model-derived deviation from expected age-related signal patterns rather than an absolute biological-age truth. Third, the HBPM cohort was cross-sectional, and the observed associations between adjusted heart age gap and prevalent hypertension do not establish causality or future incident hypertension risk. Fourth, the external PWV cohort was relatively modest in size; although it provided important vascular function validation, larger prospective and multicenter cohorts are needed to evaluate whether heart age gap predicts arterial stiffness progression, blood pressure trajectories, and clinical cardiovascular events.




\section*{METHODS}

\subsection*{Data collection and cohort construction}

Data acquisition was conducted as part of a multicenter observational study approved by the Peking University ethics review board, West China Hospital of Sichuan University, and Fuwai Hospital Chinese Academy of Medical Sciences, Shenzhen, under IRB numbers 00001052-25201, 54/2023, and SP2022112(01)A, respectively. All smartwatch physiological signals were collected using OPPO smartwatches. The study included three cohorts: the internal VHS cohort for unlabeled pretraining, supervised fine-tuning, and internal validation; the external PWV cohort for external validation and arterial stiffness association analyses; and the external HBPM cohort for external validation and hypertension association analyses. Repeated recordings in the VHS labeled partition, the PWV cohort, and the HBPM cohort were restricted to a window of no more than 3 months for each participant, to reduce the influence of long-term physiological changes on subject-level aggregation analyses.

The VHS cohort was derived from user-initiated smartwatch health monitoring and included synchronized single-lead ECG and multichannel PPG signals. The PPG channels included green-light and red-light measurements, and signals were sampled at 250 Hz. Each VHS recording lasted 30 s. Across the full VHS cohort, each participant contributed an average of 12.45 synchronized ECG--PPG recordings. The VHS cohort was not subject to strict disease-based inclusion or exclusion criteria and was primarily included according to data availability and signal quality.

The external PWV cohort was constructed from paired smartwatch PPG recordings and carotid--femoral pulse wave velocity (cfPWV) measurements obtained during the same visit. Each participant underwent smartwatch PPG acquisition and cfPWV assessment sequentially. cfPWV was measured three times, averaged, and then multiplied by 0.8 to derive the final PWV value used in the analysis. In the PWV cohort, each participant contributed an average of 7.43 smartwatch recordings, each lasting 1 min. The external HBPM cohort followed a home blood pressure monitoring design. Each participant was instructed to perform measurements during three daily time windows: morning, 06:00--08:00; midday, 12:00--14:00; and evening, 20:00--22:00. Before each measurement session, participants rested for 1 min, followed by sequential smartwatch PPG acquisition and blood pressure measurement using an Omron digital cuff sphygmomanometer. In this study, up to six PPG records per participant per day were included. Participants in the HBPM dataset contributed an average of 203.6 one-minute recordings, corresponding to approximately 34 days of repeated monitoring on average.

The external PWV and HBPM cohorts used the same major exclusion criteria to reduce the influence of known disease or treatment status on hemodynamic phenotypes. Participants were excluded if they were currently taking antihypertensive medication, had a diagnosis of pathological arrhythmia, had peripheral artery disease within the previous 6 months, or were using specific cardiovascular therapeutic devices. In contrast, the VHS dataset served as a large-scale model development dataset and did not apply these strict disease-based exclusion criteria.

\subsection*{Signal preprocessing and segment selection}

All model training, validation, and external inference were performed using 30-s signal segments. For recordings longer than 30 s, such as the 1-min recordings in the PWV and HBPM cohorts, we selected a continuous 30-s window with high signal quality according to a predefined signal-quality score. For PPG recordings with fewer than four channels, such as records from devices with only two photodiode sensor outputs, missing channels were zero-padded so that all PPG inputs had a unified four-channel format.

ECG signals were first processed using notch filtering. For each frequency in a predefined set of interference frequencies, an IIR notch filter with a quality factor of \(Q=30\) was applied for narrow-band suppression, primarily to remove power-line and related narrow-band noise. The ECG signal was then sequentially processed using high-pass and low-pass filters to remove low-frequency baseline drift and high-frequency noise. All ECG filtering steps were implemented using bidirectional zero-phase filtering to minimize phase delay and preserve ECG waveform morphology and key temporal features. The filtered ECG signal was then z-score normalized to have zero mean and unit variance. If the standard deviation of a segment was below a predefined numerical threshold, the segment was set to zero to avoid numerical instability.

PPG signals were preprocessed independently for each channel. For each PPG channel, ambient-light correction was first performed according to the device raw recording format by subtracting the raw PPG value from the ambient-light measurement, yielding the corrected photoplethysmographic pulse signal. A Butterworth band-pass filter was then applied to each PPG channel to restrict the signal to the predefined PPG frequency range, \( [f_{\mathrm{PPG,low}}, f_{\mathrm{PPG,high}}] \), preserving the dominant pulse-related frequency components while removing low-frequency drift, slow-varying trends, and high-frequency noise. The PPG band-pass filtering was also implemented using bidirectional zero-phase filtering. Finally, each PPG channel was z-score normalized independently to harmonize amplitude scales across recordings and channels. If the standard deviation of a channel segment was below the predefined numerical threshold, that channel segment was set to zero.


\subsection*{Model architecture}

All deep learning models used 30-s physiological signal segments as input. Before being fed into the model, signals were resampled to 64 Hz, yielding \(L=1920\) time points per segment. ECG input was represented as \(x^{E}\in\mathbb{R}^{1\times L}\), and PPG input was represented as \(x^{P}\in\mathbb{R}^{4\times L}\). For PPG, the four channels were provided as parallel input channels to the first one-dimensional convolutional layer, corresponding to early channel-level fusion. For the ECG+PPG fusion model, ECG and PPG were encoded by separate modality-specific encoders, and their feature embeddings were concatenated at the representation level.

Both ECG and PPG encoders used a one-dimensional ResNet-18 \cite{he2016deep} architecture. The network began with a one-dimensional convolutional stem consisting of convolution, batch normalization, ReLU activation, and max pooling, followed by four residual stages with two basic residual blocks in each stage. Each residual block contained two one-dimensional convolutional layers with batch normalization and ReLU activation, and a \(1\times1\) convolutional shortcut was used when the temporal resolution or channel width changed. The four residual stages used channel widths of 64, 128, 256, and 512. An adaptive average pooling layer generated a global temporal representation, which was mapped by a fully connected layer to a 256-dimensional encoder feature \(h\). During pretraining, this feature was further passed through a two-layer MLP projection head to obtain a 128-dimensional projected embedding \(z\) for contrastive learning. During fine-tuning, the projection head was removed, and the encoder output \(h\) was used as the input representation for heart-age regression.

\subsection*{Subject-aware self-supervised and ECG-guided cross-modal pretraining}

Pretraining was performed using unlabeled synchronized ECG--PPG recordings from the VHS cohort. For each synchronized record, ECG and PPG segments from the same time window were generated, and two augmented views were created for each modality. Augmentations included random cropping or padding, random PPG channel masking, Gaussian noise perturbation, and amplitude scaling. ECG and PPG were cropped synchronously within the same record to ensure cross-modal alignment from the same time window, whereas noise and amplitude augmentations were applied independently across modalities and views.

Single-modality SA-SSL used subject-aware contrastive learning. Specifically, a mini-batch contained \(B\) records, and two augmented views were generated for each record, yielding \(2B\) samples. For modality \(m\in\{E,P\}\), the encoder and projection head produced normalized projected embeddings \(z_i^m\). For each sample, positives were defined as other views or records from the same participant, whereas negatives were samples from other participants in the mini-batch. The subject-aware contrastive loss was defined as:
\begin{equation}
\mathcal{L}_{\mathrm{SA}}^{m}
=
-\frac{1}{2B}
\sum_{i=1}^{2B}
\frac{1}{|\mathcal{P}(i)|}
\sum_{p\in\mathcal{P}(i)}
\log
\frac{
\exp\left(\mathrm{sim}(z_i^m,z_p^m)/\tau_{\mathrm{intra}}\right)
}{
\sum_{a\neq i}
\exp\left(\mathrm{sim}(z_i^m,z_a^m)/\tau_{\mathrm{intra}}\right)
}.
\end{equation}
Here, \(\mathcal{P}(i)=\{p:p\neq i, s_p=s_i\}\) denotes the positive set containing samples from the same participant as sample \(i\), \(s_i\) is the participant identifier, \(\mathrm{sim}(\cdot,\cdot)\) denotes cosine similarity, and \(\tau_{\mathrm{intra}}\) is the temperature parameter. This objective encouraged the model to distinguish between participants while pulling together representations from different augmented views or records of the same participant, without using age labels.

SA-CLIP further introduced ECG--PPG cross-modal alignment on top of SA-SSL. For the \(i\)-th record in a mini-batch, the synchronized ECG embedding \(z_i^E\) and PPG embedding \(z_i^P\) formed a positive cross-modal pair, whereas \(z_i^E\) and \(z_j^P\) for \(j\neq i\) formed negative pairs. The cross-modal similarity matrix was defined as:
\begin{equation}
S_{ij}
=
\frac{
\mathrm{sim}(z_i^E,z_j^P)
}{
\tau_{\mathrm{xmod}}
}.
\end{equation}
The cross-modal CLIP loss used a symmetric InfoNCE objective, optimizing both ECG-to-PPG and PPG-to-ECG retrieval directions:
\begin{equation}
\mathcal{L}_{\mathrm{xmod}}
=
\frac{1}{2}
\left[
\mathrm{CE}(S,\mathbf{I})
+
\mathrm{CE}(S^{\top},\mathbf{I})
\right],
\end{equation}
where \(\mathbf{I}\) denotes the diagonal target corresponding to the correctly paired samples within the mini-batch. The final SA-CLIP pretraining objective was:
\begin{equation}
\mathcal{L}_{\mathrm{SA\text{-}CLIP}}
=
\lambda_P \mathcal{L}_{\mathrm{SA}}^{P}
+
\lambda_E \mathcal{L}_{\mathrm{SA}}^{E}
+
\lambda_X \mathcal{L}_{\mathrm{xmod}}.
\end{equation}

In this study, \(\lambda_P=1.0\), \(\lambda_E=1.0\), \(\lambda_X=0.5\), \(\tau_{\mathrm{intra}}=0.1\), and \(\tau_{\mathrm{xmod}}=0.07\). Pretraining used the AdamW optimizer with a learning rate of \(3\times10^{-4}\), weight decay of \(1\times10^{-4}\), and 100 epochs. The batch size was 2048 for single-modality SA-SSL and 256 for multimodal SA-CLIP, with cosine annealing used for the multimodal pretraining schedule. After pretraining, ECG and PPG encoder weights were saved separately for downstream heart-age prediction.

\subsection*{Supervised fine-tuning for heart-age prediction}

The heart-age prediction task was fine-tuned on age-labeled VHS data using chronological age as the regression target. We used a subject-level split to avoid leakage across records from the same participant. Specifically, participants were randomly divided into training, validation, and test sets at a ratio of 7:1:2. The model was optimized on the training set, selected according to validation MAE, and finally evaluated on the held-out test set for internal validation. The external PWV and HBPM cohorts were not used for training or model selection and were reserved for external inference and downstream association analyses.

During fine-tuning, the pretrained ECG and PPG encoders were loaded, and the projection heads were removed. For PPG-only or ECG-only models, the 256-dimensional encoder feature was passed to a linear regression head to predict heart age:
\begin{equation}
\widehat{y}_i = \mathbf{w}^{\top}h_i + b.
\end{equation}
For the ECG+PPG fusion model, ECG and PPG encoders produced \(h_i^E\) and \(h_i^P\), which were concatenated and passed to a linear regression head:
\begin{equation}
\widehat{y}_i = \mathbf{w}^{\top}[h_i^E;h_i^P] + b.
\end{equation}
The primary supervised fine-tuning loss was mean squared error:
\begin{equation}
\mathcal{L}_{\mathrm{age}}
=
\frac{1}{N}
\sum_{i=1}^{N}
(\widehat{y}_i-y_i)^2,
\end{equation}
where \(y_i\) denotes the chronological age at the time of recording. Fine-tuning used AdamW with a learning rate of \(3\times10^{-4}\), weight decay of \(1\times10^{-4}\), batch size of 512, and a maximum of 50 epochs. Early stopping was based on validation MAE with a patience of 7 epochs. Unless otherwise specified, encoder parameters were updated end-to-end together with the regression head during fine-tuning.

Model performance was evaluated at both the record and subject levels. Record-level metrics were calculated directly from predictions for individual 30-s records. Subject-level prediction was obtained by averaging all available record-level predictions from the same participant:
\begin{equation}
\widehat{y}_s
=
\frac{1}{|\mathcal{R}_s|}
\sum_{i\in\mathcal{R}_s}
\widehat{y}_i,
\end{equation}
where \(\mathcal{R}_s\) denotes the set of records from participant \(s\). Subject-level heart age gap was defined as:
\begin{equation}
\Delta_{\mathrm{HA},s}
=
\widehat{y}_s-y_s.
\end{equation}

For model comparison, ECG-only, PPG-only, and ECG+PPG fusion models were evaluated. For external validation and subsequent PWV and hypertension association analyses, the SA-CLIP-pretrained PPG-only model was used as the primary deployment model, matching the intended smartwatch PPG-only use case.


\subsection*{Temporal aggregation and scaling-law analysis}

To evaluate the effect of short-term repeated-record aggregation on the stability of subject-level heart-age estimation, we performed two scaling-law analyses using record-level predictions from the internal VHS test set. Because all repeated recordings included in this analysis were acquired within a short-term window of no more than 3 months for each participant, we assumed that participants remained in a relatively stable physiological state during this period and that their underlying heart-age phenotype was approximately stable. The first analysis assessed the aggregation effect as the number of available records per participant increased. Specifically, \(k\) was varied from 1 to 15; for each value of \(k\), up to \(k\) available records per participant were included, and their record-level predicted ages were averaged to obtain a subject-level prediction. MAE, RMSE, Pearson's \(r\), and \(R^2\) were then calculated at the subject level. For participants with fewer than \(k\) available records, all available records were used. The second analysis evaluated the effect of per-participant data density. The minimum-record threshold \(N_{\min}\) was varied from 1 to 15; only participants with at least \(N_{\min}\) records were retained, all available record-level predictions from each retained participant were averaged, and subject-level performance metrics and the number of remaining participants were calculated. Together, these two analyses were used to assess the potential value of short-term repeated measurements for improving the accuracy and stability of smartwatch-based subject-level heart-age estimation.


\subsection*{Statistical analysis}

Unless otherwise specified, all downstream statistical analyses were performed at the subject level. Record-level predictions from the same participant were first aggregated to obtain subject-level predicted heart age, and heart age gap was calculated as the difference between predicted heart age and chronological age. In the external PWV cohort, partial correlation analysis was first used to evaluate the association between heart age gap and PWV independent of chronological age. Specifically, the linear effects of chronological age were removed from both heart age gap and PWV, and Pearson correlation was then calculated between the two residualized variables. We then fitted multivariable linear regression models with PWV as the continuous outcome and heart age gap as the primary exposure, adjusting for chronological age, sex, and BMI. Linear regression models used heteroscedasticity-robust standard errors, and results were reported as unstandardized regression coefficients, 95\% confidence intervals, and two-sided \(P\) values. For PWV-related analyses, continuous variables were winsorized at the 1st and 99th percentiles when appropriate to reduce the influence of extreme values on model estimation.

In the external HBPM cohort, to reduce systematic age-related bias in the raw heart age gap, we fitted a generalized additive model with chronological age as a smooth term and defined the model residual as the adjusted heart age gap. In descriptive stratified analyses, participants were grouped by chronological age into \(<30\), \([30,40)\), \([40,50)\), and \([50,60)\) years, and by adjusted heart age gap into \(\leq0\), \((0,2]\), \((2,4]\), and \(>4\) years. Hypertension prevalence and Wilson score 95\% confidence intervals were calculated within each age-by-gap stratum, and relative prevalence was defined as the prevalence within each gap stratum divided by the average prevalence within the same chronological-age group. We further fitted multivariable logistic regression models with prevalent hypertension as the binary outcome and adjusted heart age gap modeled both as a standardized continuous variable and as quartiles. The overall model was adjusted for chronological age, sex, BMI, and family history of hypertension; sex-stratified models were adjusted for chronological age, BMI, and family history of hypertension. Logistic regression results were reported as odds ratios, 95\% confidence intervals, and two-sided \(P\) values. All statistical tests were two-sided, with statistical significance defined as \(P<0.05\). Because the HBPM cohort was cross-sectional, hypertension analyses were interpreted as associations with prevalent hypertension rather than prospective incidence risk.

\newpage

\section*{RESOURCE AVAILABILITY}
\subsection*{Data availability}
The cohorts analyzed in this study were collected and are owned by OPPO Health Lab as
part of a multicenter observational study and contain sensitive commercial and participant
health information. As such, the raw data are not publicly available at this time. Access to
de-identified data may be considered for collaborative research upon reasonable request and
subject to approval by OPPO Health Lab and the relevant institutional review boards.

\subsection*{Code availability}
The code used for model pretraining, fine-tuning, and downstream statistical
analyses contains proprietary components developed as part of an ongoing
industry-academia collaboration and is not publicly available at this time.
Requests for access to specific code components for the purpose of result
verification or academic collaboration may be directed to the corresponding
author and will be considered on a case-by-case basis, subject to approval
by OPPO Health Lab.

\section*{ACKNOWLEDGMENTS}


This work was supported by the PKU-OPPO Fund (BO202504).

\section*{AUTHOR CONTRIBUTIONS}


D.X. led the overall algorithm design and implementation, and drafted the manuscript.
X.G., Y.Z. and F.X. led early-stage data collection and preliminary data organization.
G.N., C.X., J.L. and S.T. contributed to data processing and provided suggestions on
algorithmic improvements. X.L. and Q.X. contributed to manuscript revision. Y.L. and S.H.
supervised the project and contributed to the overall study design. All authors reviewed and
approved the final manuscript.

\section*{DECLARATION OF INTERESTS}


The authors declare no competing interests.

\newpage


\bibliography{reference}

\bigskip


\newpage
\begin{appendices}

\section{Supplementary Tables for Short-Term Multi-Record Aggregation}
\label{app:scaling_law}

\subsection{Scaling-Law Analysis of Heart-Age Estimation with Increasing Aggregated Recordings}
\begin{table}[h]
\centering
\caption{
\textbf{Scaling-law analysis of heart-age estimation with short-term multi-record aggregation.} Performance was evaluated at the subject level in the internal VHS cohort after averaging predictions from $k$ recordings per participant.
MAE and RMSE are reported in years.
Pearson's \textit{r} denotes the correlation between predicted heart age and chronological age.
$R^2$ denotes the coefficient of determination.
All repeated recordings were acquired within a short-term window of no more than 3 months.
}
\label{tab:scaling_law_multisample}
\small
\setlength{\tabcolsep}{16pt}
\renewcommand{\arraystretch}{1.12}

\begin{tabular}{ccccc}
\toprule
\makecell{\textbf{Number of recordings}\\\textbf{per participant} ($k$)} &
\textbf{MAE $\downarrow$} &
\textbf{RMSE $\downarrow$} &
\textbf{Pearson's \textit{r} $\uparrow$} &
\textbf{$R^2$ $\uparrow$} \\
\midrule
1  & 5.395 & 7.069 & 0.649 & 0.414 \\
2  & 5.143 & 6.689 & 0.692 & 0.475 \\
3  & 5.065 & 6.584 & 0.706 & 0.492 \\
4  & 5.032 & 6.523 & 0.714 & 0.501 \\
5  & 5.023 & 6.506 & 0.717 & 0.504 \\
6  & 4.998 & 6.466 & 0.723 & 0.510 \\
7  & 4.980 & 6.441 & 0.726 & 0.514 \\
8  & 4.972 & 6.427 & 0.728 & 0.516 \\
9  & 4.961 & 6.414 & 0.730 & 0.518 \\
10 & 4.963 & 6.413 & 0.730 & 0.518 \\
11 & 4.955 & 6.406 & 0.731 & 0.519 \\
12 & 4.951 & 6.399 & 0.732 & 0.520 \\
13 & 4.950 & 6.398 & 0.732 & 0.520 \\
14 & 4.950 & 6.396 & 0.733 & 0.520 \\
15 & 4.949 & 6.394 & 0.733 & 0.521 \\
\bottomrule
\end{tabular}

\end{table}

\newpage
\subsection{Subject-Level Prediction Performance by Minimum Recording Density Threshold}
\begin{table}[h]
\centering
\caption{
\textbf{Subject-level prediction performance after filtering participants by minimum recording density.} Participants were retained only if they had at least $N_{\min}$ recordings.
Subject-level performance was calculated after aggregating predictions within each retained participant.
MAE and RMSE are reported in years.
Pearson's \textit{r} denotes the correlation between predicted heart age and chronological age.
$R^2$ denotes the coefficient of determination.
}
\label{tab:threshold_density}
\small
\setlength{\tabcolsep}{7pt}
\renewcommand{\arraystretch}{1.12}
\resizebox{\textwidth}{!}{
\begin{tabular}{cccccc}
\toprule
\makecell{\textbf{Minimum recordings}\\\textbf{per participant} ($N_{\min}$)} &
\makecell{\textbf{Remaining}\\\textbf{participants}} &
\textbf{MAE $\downarrow$} &
\textbf{RMSE $\downarrow$} &
\textbf{Pearson's \textit{r} $\uparrow$} &
\textbf{$R^2$ $\uparrow$} \\
\midrule
1  & 8773 & 4.951 & 6.394 & 0.734 & 0.521 \\
2  & 7431 & 4.874 & 6.280 & 0.761 & 0.547 \\
3  & 6555 & 4.848 & 6.221 & 0.775 & 0.556 \\
4  & 5899 & 4.828 & 6.180 & 0.782 & 0.563 \\
5  & 5316 & 4.825 & 6.144 & 0.790 & 0.571 \\
6  & 4828 & 4.829 & 6.157 & 0.795 & 0.574 \\
7  & 4416 & 4.836 & 6.132 & 0.801 & 0.579 \\
8  & 4055 & 4.845 & 6.104 & 0.805 & 0.583 \\
9  & 3749 & 4.878 & 6.141 & 0.806 & 0.583 \\
10 & 3463 & 4.930 & 6.198 & 0.806 & 0.582 \\
11 & 3214 & 4.943 & 6.232 & 0.804 & 0.579 \\
12 & 3002 & 4.923 & 6.196 & 0.809 & 0.586 \\
13 & 2787 & 4.937 & 6.223 & 0.811 & 0.588 \\
14 & 2624 & 4.954 & 6.235 & 0.813 & 0.589 \\
15 & 2447 & 4.963 & 6.245 & 0.817 & 0.592 \\
\bottomrule
\end{tabular}
}
\vspace{2mm}
\begin{minipage}{0.98\textwidth}
\footnotesize
\end{minipage}
\end{table}


\clearpage

\begin{landscape}
\setlength{\LTcapwidth}{\linewidth}
\begin{longtable}{
  >{\raggedright\arraybackslash}p{3.0cm}
  >{\centering\arraybackslash}p{1.8cm}
  >{\centering\arraybackslash}p{2.8cm}
  >{\centering\arraybackslash}p{1.0cm}
  >{\centering\arraybackslash}p{2.2cm}
  >{\centering\arraybackslash}p{3.5cm}
  >{\centering\arraybackslash}p{4.0cm}
}

\caption{Detailed hypertension prevalence and relative prevalence across
chronological-age and adjusted heart age gap strata.}
\label{tab:supp_hypertension_gap} \\

\toprule
\textbf{Group} &
\textbf{Age group} &
\textbf{Adjusted heart age gap} &
\textbf{N} &
\textbf{Hypertension cases} &
\textbf{Prevalence, \% (95\% CI)} &
\textbf{Relative prevalence (95\% CI)} \\
\midrule
\endfirsthead

\multicolumn{7}{l}{\small\textit{Table~\ref{tab:supp_hypertension_gap} continued}} \\[2pt]
\toprule
\textbf{Group} &
\textbf{Age group} &
\textbf{Adjusted heart age gap} &
\textbf{N} &
\textbf{Hypertension cases} &
\textbf{Prevalence, \% (95\% CI)} &
\textbf{Relative prevalence (95\% CI)} \\
\midrule
\endhead

\midrule
\multicolumn{7}{r}{\small\textit{Continued on next page}} \\
\endfoot

\bottomrule
\multicolumn{7}{p{20cm}}{%
  \footnotesize\textit{Note.} Relative prevalence was defined as the
  hypertension prevalence within each adjusted heart age gap stratum
  divided by the average prevalence within the same chronological-age
  group. Confidence intervals were estimated using bootstrap resampling.
  Because the HBPM cohort is cross-sectional, these values should be
  interpreted as prevalence and relative prevalence rather than
  prospective incidence risk. Age-by-gap strata with fewer than 5
  participants were omitted from this table to avoid reporting unstable
  estimates; consequently, the sum of $N$ across strata is smaller than
  the total HBPM cohort size (1,095 participants).} \\
\endlastfoot

All participants & $<30$     & $\leq 0$  & 167 & 12 & 7.2 (4.2--12.1)    & 0.56 (0.32--0.95) \\
All participants & $<30$     & $(0,2]$   & 32  & 4  & 12.5 (5.0--28.1)   & 0.98 (0.39--2.19) \\
All participants & $<30$     & $(2,4]$   & 10  & 2  & 20.0 (5.7--51.0)   & 1.56 (0.44--3.98) \\
All participants & $<30$     & $>4$      & 25  & 12 & 48.0 (30.0--66.5)  & 3.74 (2.34--5.19) \\[4pt]

All participants & $[30,40)$ & $\leq 0$  & 238 & 35 & 14.7 (10.8--19.8)  & 0.58 (0.42--0.78) \\
All participants & $[30,40)$ & $(0,2]$   & 103 & 33 & 32.0 (23.8--41.6)  & 1.26 (0.94--1.64) \\
All participants & $[30,40)$ & $(2,4]$   & 35  & 19 & 54.3 (38.2--69.5)  & 2.14 (1.51--2.74) \\
All participants & $[30,40)$ & $>4$      & 46  & 20 & 43.5 (30.2--57.8)  & 1.71 (1.19--2.28) \\[4pt]

All participants & $[40,50)$ & $\leq 0$  & 166 & 54 & 32.5 (25.9--40.0)  & 0.76 (0.60--0.93) \\
All participants & $[40,50)$ & $(0,2]$   & 49  & 25 & 51.0 (37.5--64.4)  & 1.19 (0.87--1.50) \\
All participants & $[40,50)$ & $(2,4]$   & 50  & 26 & 52.0 (38.5--65.2)  & 1.21 (0.90--1.52) \\
All participants & $[40,50)$ & $>4$      & 50  & 30 & 60.0 (46.2--72.4)  & 1.40 (1.08--1.69) \\[4pt]

All participants & $[50,60)$ & $\leq 0$  & 55  & 28 & 50.9 (38.1--63.6)  & 0.87 (0.65--1.09) \\
All participants & $[50,60)$ & $(0,2]$   & 15  & 6  & 40.0 (19.8--64.3)  & 0.69 (0.34--1.10) \\
All participants & $[50,60)$ & $(2,4]$   & 23  & 18 & 78.3 (58.1--90.3)  & 1.34 (1.00--1.55) \\
All participants & $[50,60)$ & $>4$      & 22  & 15 & 68.2 (47.3--83.6)  & 1.17 (0.81--1.44) \\[8pt]

Female & $<30$     & $\leq 0$  & 69  & 3  & 4.3 (1.5--12.0)    & 0.33 (0.11--0.92) \\
Female & $<30$     & $(0,2]$   & 24  & 3  & 12.5 (4.3--31.0)   & 0.96 (0.33--2.38) \\
Female & $<30$     & $(2,4]$   & 7   & 2  & 28.6 (8.2--64.1)   & 2.19 (0.63--4.91) \\
Female & $<30$     & $>4$      & 15  & 7  & 46.7 (24.8--69.9)  & 3.58 (1.90--5.36) \\[4pt]

Female & $[30,40)$ & $\leq 0$  & 105 & 13 & 12.4 (7.4--20.0)   & 0.48 (0.29--0.78) \\
Female & $[30,40)$ & $(0,2]$   & 48  & 15 & 31.2 (19.9--45.3)  & 1.21 (0.78--1.76) \\
Female & $[30,40)$ & $(2,4]$   & 23  & 11 & 47.8 (29.2--67.0)  & 1.86 (1.14--2.61) \\
Female & $[30,40)$ & $>4$      & 30  & 14 & 46.7 (30.2--63.9)  & 1.81 (1.18--2.48) \\[4pt]

Female & $[40,50)$ & $\leq 0$  & 68  & 10 & 14.7 (8.2--25.0)   & 0.42 (0.23--0.71) \\
Female & $[40,50)$ & $(0,2]$   & 20  & 8  & 40.0 (21.9--61.3)  & 1.13 (0.62--1.74) \\
Female & $[40,50)$ & $(2,4]$   & 34  & 16 & 47.1 (31.5--63.3)  & 1.33 (0.89--1.79) \\
Female & $[40,50)$ & $>4$      & 34  & 21 & 61.8 (45.0--76.1)  & 1.75 (1.28--2.16) \\[4pt]

Female & $[50,60)$ & $\leq 0$  & 26  & 11 & 42.3 (25.5--61.1)  & 0.77 (0.47--1.12) \\
Female & $[50,60)$ & $(0,2]$   & 9   & 2  & 22.2 (6.3--54.7)   & 0.41 (0.12--1.00) \\
Female & $[50,60)$ & $(2,4]$   & 16  & 12 & 75.0 (50.5--89.8)  & 1.37 (0.92--1.64) \\
Female & $[50,60)$ & $>4$      & 13  & 10 & 76.9 (49.7--91.8)  & 1.41 (0.91--1.68) \\[8pt]

Male & $<30$     & $\leq 0$  & 98  & 9  & 9.2 (4.9--16.5)    & 0.73 (0.39--1.31) \\
Male & $<30$     & $(0,2]$   & 8   & 1  & 12.5 (2.2--47.1)   & 0.99 (0.18--3.74) \\
Male & $<30$     & $(2,4]$   & 3   & 0  & 0.0 (0.0--56.1)    & 0.00 (0.00--4.45) \\
Male & $<30$     & $>4$      & 10  & 5  & 50.0 (23.7--76.3)  & 3.97 (1.88--6.06) \\[4pt]

Male & $[30,40)$ & $\leq 0$  & 133 & 22 & 16.5 (11.2--23.8)  & 0.66 (0.45--0.95) \\
Male & $[30,40)$ & $(0,2]$   & 55  & 18 & 32.7 (21.8--45.9)  & 1.31 (0.87--1.84) \\
Male & $[30,40)$ & $(2,4]$   & 12  & 8  & 66.7 (39.1--86.2)  & 2.67 (1.56--3.45) \\
Male & $[30,40)$ & $>4$      & 16  & 6  & 37.5 (18.5--61.4)  & 1.50 (0.74--2.45) \\[4pt]

Male & $[40,50)$ & $\leq 0$  & 98  & 44 & 44.9 (35.4--54.8)  & 0.89 (0.70--1.09) \\
Male & $[40,50)$ & $(0,2]$   & 29  & 17 & 58.6 (40.7--74.5)  & 1.17 (0.81--1.48) \\
Male & $[40,50)$ & $(2,4]$   & 16  & 10 & 62.5 (38.6--81.5)  & 1.24 (0.77--1.62) \\
Male & $[40,50)$ & $>4$      & 16  & 9  & 56.2 (33.2--76.9)  & 1.12 (0.66--1.53) \\[4pt]

Male & $[50,60)$ & $\leq 0$  & 29  & 17 & 58.6 (40.7--74.5)  & 0.93 (0.65--1.19) \\
Male & $[50,60)$ & $(0,2]$   & 6   & 4  & 66.7 (30.0--90.3)  & 1.06 (0.48--1.44) \\
Male & $[50,60)$ & $(2,4]$   & 7   & 6  & 85.7 (48.7--97.4)  & 1.37 (0.78--1.55) \\
Male & $[50,60)$ & $>4$      & 9   & 5  & 55.6 (26.7--81.1)  & 0.89 (0.42--1.29) \\

\end{longtable}
\end{landscape}

\clearpage

\begin{landscape}
\setlength{\LTcapwidth}{\linewidth}
\begin{longtable}{
  >{\raggedright\arraybackslash}p{3.0cm}
  >{\centering\arraybackslash}p{2.8cm}
  >{\centering\arraybackslash}p{2.8cm}
  >{\centering\arraybackslash}p{1.0cm}
  >{\centering\arraybackslash}p{4.5cm}
  >{\centering\arraybackslash}p{4.5cm}
}

\caption{Hypertension prevalence across chronological-age groups and
adjusted heart age gap quartiles, stratified by sex.}
\label{tab:supp_hypertension_quartile} \\

\toprule
\textbf{Group} &
\textbf{Age group} &
\textbf{Gap quartile} &
\textbf{N} &
\textbf{Hypertension cases} &
\textbf{Prevalence (95\% CI)} \\
\midrule
\endfirsthead

\multicolumn{6}{l}{\small\textit{Table~\ref{tab:supp_hypertension_quartile} continued}} \\[2pt]
\toprule
\textbf{Group} &
\textbf{Age group} &
\textbf{Gap quartile} &
\textbf{N} &
\textbf{Hypertension cases} &
\textbf{Prevalence (95\% CI)} \\
\midrule
\endhead

\midrule
\multicolumn{6}{r}{\small\textit{Continued on next page}} \\
\endfoot

\bottomrule
\multicolumn{6}{p{20cm}}{%
  \footnotesize\textit{Note.}
  Gap quartiles were defined based on the adjusted heart age gap
  $\Delta_{\text{HA,adj}}$, with Q1 (lowest) to Q4 (highest) representing
  increasing levels of adjusted heart age gap.
  Prevalence was estimated within each age-by-quartile stratum,
  with 95\% confidence intervals computed using the Wilson score method.
  These values are provided to contextualize the multivariable logistic
  regression results reported in the main text.
  Because the HBPM cohort is cross-sectional, estimates represent
  observed prevalence of hypertension rather than prospective incidence risk.
  Age-by-quartile strata with fewer than 5 participants were omitted from
  this table to avoid reporting unstable estimates; consequently, the sum
  of $N$ across strata is smaller than the total HBPM cohort size
  (1,095 participants).
} \\
\endlastfoot

All participants & $<30$     & Q1 (lowest)  & 54  & 4  & 7.4\% (2.9\%--17.6\%)  \\
All participants & $<30$     & Q2           & 94  & 3  & 3.2\% (1.1\%--9.0\%)   \\
All participants & $<30$     & Q3           & 51  & 9  & 17.6\% (9.6\%--30.3\%) \\
All participants & $<30$     & Q4 (highest) & 35  & 14 & 40.0\% (25.6\%--56.4\%)\\[4pt]

All participants & $[30,40)$ & Q1 (lowest)  & 92  & 11 & 12.0\% (6.8\%--20.2\%) \\
All participants & $[30,40)$ & Q2           & 107 & 17 & 15.9\% (10.2\%--24.0\%)\\
All participants & $[30,40)$ & Q3           & 135 & 37 & 27.4\% (20.6\%--35.5\%)\\
All participants & $[30,40)$ & Q4 (highest) & 88  & 42 & 47.7\% (37.6\%--58.0\%)\\[4pt]

All participants & $[40,50)$ & Q1 (lowest)  & 94  & 27 & 28.7\% (20.6\%--38.6\%)\\
All participants & $[40,50)$ & Q2           & 52  & 15 & 28.8\% (18.3\%--42.3\%)\\
All participants & $[40,50)$ & Q3           & 65  & 36 & 55.4\% (43.3\%--66.8\%)\\
All participants & $[40,50)$ & Q4 (highest) & 104 & 57 & 54.8\% (45.2\%--64.0\%)\\[4pt]

All participants & $[50,60)$ & Q1 (lowest)  & 31  & 14 & 45.2\% (29.2\%--62.2\%)\\
All participants & $[50,60)$ & Q2           & 19  & 11 & 57.9\% (36.3\%--76.9\%)\\
All participants & $[50,60)$ & Q3           & 20  & 9  & 45.0\% (25.8\%--65.8\%)\\
All participants & $[50,60)$ & Q4 (highest) & 45  & 33 & 73.3\% (59.0\%--84.0\%)\\[8pt]

Male & $<30$     & Q1 (lowest)  & 35 & 3  & 8.6\% (3.0\%--22.4\%)  \\
Male & $<30$     & Q2           & 50 & 3  & 6.0\% (2.1\%--16.2\%)  \\
Male & $<30$     & Q3           & 21 & 4  & 19.0\% (7.7\%--40.0\%) \\
Male & $<30$     & Q4 (highest) & 13 & 5  & 38.5\% (17.7\%--64.5\%)\\[4pt]

Male & $[30,40)$ & Q1 (lowest)  & 58 & 6  & 10.3\% (4.8\%--20.8\%) \\
Male & $[30,40)$ & Q2           & 64 & 12 & 18.8\% (11.1\%--30.0\%)\\
Male & $[30,40)$ & Q3           & 62 & 20 & 32.3\% (22.0\%--44.6\%)\\
Male & $[30,40)$ & Q4 (highest) & 32 & 16 & 50.0\% (33.6\%--66.4\%)\\[4pt]

Male & $[40,50)$ & Q1 (lowest)  & 54 & 22 & 40.7\% (28.7\%--54.0\%)\\
Male & $[40,50)$ & Q2           & 32 & 12 & 37.5\% (22.9\%--54.7\%)\\
Male & $[40,50)$ & Q3           & 39 & 27 & 69.2\% (53.6\%--81.4\%)\\
Male & $[40,50)$ & Q4 (highest) & 34 & 19 & 55.9\% (39.5\%--71.1\%)\\[4pt]

Male & $[50,60)$ & Q1 (lowest)  & 19 & 10 & 52.6\% (31.7\%--72.7\%)\\
Male & $[50,60)$ & Q2           & 9  & 6  & 66.7\% (35.4\%--87.9\%)\\
Male & $[50,60)$ & Q3           & 7  & 5  & 71.4\% (35.9\%--91.8\%)\\
Male & $[50,60)$ & Q4 (highest) & 16 & 11 & 68.8\% (44.4\%--85.8\%)\\[8pt]

Female & $<30$     & Q1 (lowest)  & 19 & 1  & 5.3\% (0.9\%--24.6\%)  \\
Female & $<30$     & Q2           & 44 & 0  & 0.0\% (0.0\%--8.0\%)   \\
Female & $<30$     & Q3           & 30 & 5  & 16.7\% (7.3\%--33.6\%) \\
Female & $<30$     & Q4 (highest) & 22 & 9  & 40.9\% (23.3\%--61.3\%)\\[4pt]

Female & $[30,40)$ & Q1 (lowest)  & 34 & 5  & 14.7\% (6.4\%--30.1\%) \\
Female & $[30,40)$ & Q2           & 43 & 5  & 11.6\% (5.1\%--24.5\%) \\
Female & $[30,40)$ & Q3           & 73 & 17 & 23.3\% (15.1\%--34.2\%)\\
Female & $[30,40)$ & Q4 (highest) & 56 & 26 & 46.4\% (34.0\%--59.3\%)\\[4pt]

Female & $[40,50)$ & Q1 (lowest)  & 40 & 5  & 12.5\% (5.5\%--26.1\%) \\
Female & $[40,50)$ & Q2           & 20 & 3  & 15.0\% (5.2\%--36.0\%) \\
Female & $[40,50)$ & Q3           & 26 & 9  & 34.6\% (19.4\%--53.8\%)\\
Female & $[40,50)$ & Q4 (highest) & 70 & 38 & 54.3\% (42.7\%--65.4\%)\\[4pt]

Female & $[50,60)$ & Q1 (lowest)  & 12 & 4  & 33.3\% (13.8\%--60.9\%)\\
Female & $[50,60)$ & Q2           & 10 & 5  & 50.0\% (23.7\%--76.3\%)\\
Female & $[50,60)$ & Q3           & 13 & 4  & 30.8\% (12.7\%--57.6\%)\\
Female & $[50,60)$ & Q4 (highest) & 29 & 22 & 75.9\% (57.9\%--87.8\%)\\

\end{longtable}
\end{landscape}

\newpage

\end{appendices}

\end{document}